\documentclass[iopams]{iopart}
\usepackage{geometry}
\usepackage{graphics}
\usepackage{graphicx}
\usepackage{amssymb}
\usepackage{enumerate}
\usepackage{color}
\usepackage{subfig}

\DeclareMathAlphabet\mathbit
    \encodingdefault\rmdefault\bfdefault\itdefault
\DeclareOldFontCommand{\bi}{\normalfont\bfseries\itshape}{\mathbit}

\newcommand{\be}{\begin{equation}}
\newcommand{\ee}{\end{equation}}

\def\fakebold#1{\relax\ifvmode\leavevmode\fi%
\ifmmode%
\setbox0=\hbox{$#1$}%
\else%
\setbox0=\hbox{#1}%
\fi%
\kern-.02em\copy0 \kern-\wd0%
\kern .04em\copy0 \kern-\wd0%
\kern-.0125em\raise.02em\box0%
}%

\begin{document}
\ifx\aligned\undefined

\makeatletter
\def\aligned{{\ifnum0=`}\fi\vcenter\bgroup\let\\\cr
\halign\bgroup&\hfil$\displaystyle{}##{}$&$\displaystyle{}##{}$\hfil\cr}
\def\endaligned{\crcr\egroup\egroup\ifnum0=`{\fi}}

\def\align{\par
\bigskip
{\ifnum0=`}\fi
\let\\\cr
\halign to \textwidth\bgroup
\refstepcounter{equation}%
\global\let\@alignlab\@currentlabel
\vrule \@height \dimexpr\ht\strutbox+3pt\relax
       \@depth  \dimexpr\dp\strutbox+1pt\relax
       \@width \z@
\hbox to \textwidth{\hfill(\theequation)}\kern-\textwidth
\tabskip\fill
\hfil$\displaystyle{}##{}$&%
\let\@currentlabel\@alignlab$\displaystyle{}##{}$\hfil&%
\let\@currentlabel\@alignlab\hfil$\displaystyle{}##{}$&%
\let\@currentlabel\@alignlab$\displaystyle{}##{}$\hfil\cr}
\def\endalign{\crcr\egroup\ifnum0=`{\fi}\par\bigskip}

\newcounter{parentequation}
\newenvironment{subequations}{%
  \refstepcounter{equation}%
  \edef\theparentequation{\theequation}%
  \setcounter{parentequation}{\value{equation}}%
  \setcounter{equation}{0}%
  \def\theequation{\theparentequation\alph{equation}}%
  \ignorespaces
}{%
  \setcounter{equation}{\value{parentequation}}%
  \ignorespacesafterend
}

\def\cases{{\ifnum0=`}\fi\left\{\array{lll}}
\def\endcases{\endarray\right.\ifnum0=`{\fi}}

\makeatother
\fi

\title[stability~of~classically~neutral~flows]{On the stability of waves in classically neutral flows}
\author{Colin Huber$^1$, Meaghan Hoitt$^1$, Nathaniel S. Barlow$^{1}$, Nicole Hill$^{2}$, Kimberlee Keithley$^{1,2}$, Steven J. Weinstein$^{1,2}$}
\address{$^1$ Department of Chemical Engineering, Rochester Institute of Technology, Rochester, NY 14623} 
\address{$^2$ School of Mathematical Sciences, Rochester Institute of Technology, Rochester, NY 14623} 
\ead{nsbsma@rit.edu}

\begin{abstract}
This paper reports a breakdown in linear stability theory under conditions of neutral stability that is deduced by an examination of exponential modes of the form $h\approx {{e}^{i(kx-\omega t)}}$, where $h$ is a response to a disturbance, $k$ is a real wavenumber, and $\omega(k)$ is a wavelength-dependent complex frequency.  In a previous paper, King et al (Stability of algebraically unstable dispersive flows, \textit{Phys. Rev. Fluids}, 1(073604), 2016) demonstrates that when Im$[\omega(k)]$=0 for all $k$, it is possible for a system response to grow or damp algebraically as $h\approx {{t}^{s}}$ where $s$ is a fractional power.  The growth is deduced through an asymptotic analysis of the Fourier integral that inherently invokes the superposition of an infinite number of modes.  In this paper, the more typical case associated with the transition from stability to instability is examined in which Im$[\omega(k)]$=0 for a single mode (i.e., for one value of $k$) at neutral stability.  Two partial differential equation systems are examined, one that has been constructed to elucidate key features of the stability threshold, and a second that models the well-studied problem of rectilinear Newtonian flow down an inclined plane.  In both cases, algebraic growth/decay is deduced at the neutral stability boundary, and the propagation features of the responses are examined.
\end{abstract}



\section{Introduction}
The current work is motivated by the need to characterize fluid flows involved in the manufacture of a variety of products. All flow processes are subjected to time-varying disturbances induced from their surroundings, while product quality typically necessitates the requirement of time invariance. In particular, each product has a manufacturing tolerance to perturbations that requires accurate characterization. Mathematical models that relate disturbances to system responses are commonly used in industry to enable guided experiments, and once validated, can be accurate enough to replace experiments. An essential feature of disturbance modeling is whether the underlying fluid system is stable or unstable, i.e. whether a response to disturbances grows or decays. If the fluid system is stable, then product specifications may be met by minimizing the magnitude of process disturbances; if the system is not, control of the process is much more complex, and in some cases, impossible. The characterization of fluid flow stability then, is motivated by practical need. Within stable fluid systems, tight product tolerances often dictate that perturbations be small and lie in a linear regime -- a product often becomes unsalable well within the dictates of linearity. Thus, a corresponding linear operator may be used to model the fluid system response. The characterization of the stability of such a linear operator, then, is of paramount importance to develop sustainable operating parameters that meet product specifications.

If a flowing liquid system in a manufacturing process is unstable, i.e., initiated disturbances are magnified, the flow can often easily be disrupted away from a uniform state. This could be a desired outcome, as is the case for liquid fuel atomizers~\cite{ibrahim1995,lin2003,el2015} where instability leads to the breakup of a liquid sheet into droplets, or in the case of instability-driven turbulence to enhance mixing processes~\cite{paul2004}. Instability is unwelcome in other situations where layer uniformity is essential, such as in the thin films used to coat ink jet and copier papers, printed electronics, and liquid crystal display screens~\cite{cohen1992,kistler1997}. Thus, it is important for practitioners to control the parameters that influence fluid instability in order to produce a salable product. The most widely used stability assessment, referred to here as classical stability theory, provides the basis for much of the hydrodynamic literature~\cite{Chandrasekhar,HuerreRossi,huerre2000,schmid2001,drazin2004}.

Classical stability theory is built on the following ideas. Typically, the analysis starts with a full nonlinear operator and boundary conditions, which are linearized about a base fluid flow (oftentimes exact) whose stability is to be assessed.  The resulting linear partial differential equation (PDE) system may be expressed in terms of an operator, $L$, and flow response, $h(x,t)$, as $Lh=0$, where $x$ and $t$ are respective space and time variables.  Note that forcing and initial conditions are neglected when examining the PDE system, as the general homogeneous response characterizes the classical stability of the medium. A solution of this equation on an unbounded domain ($-\infty < x < \infty$) may be expressed as
\begin{equation}
h=C{{e}^{i(kx-\omega t)}},
\label{eq:mode}
\end{equation}
where $k$ is a real wavenumber, $\omega$ is a complex frequency, a $C$ is an amplitude.  It is assumed that the fundamental responses~(\ref{eq:mode}) for each value of $k$, henceforth called modes, may be superimposed to build the flow response to any disturbance; furthermore, it is assumed that the stability of a complex flow response may be characterized by the stability of its constituent modes.  Substitution of~(\ref{eq:mode}) into the linearized PDE system leads to a dispersion relation of the form
\begin{equation}
D(k,\omega)=0,~\textrm{ also written }\omega=\omega(k),
\label{eq:generic}
\end{equation}
that assures a nontrivial solution of the equation $Lh=0$.  In classical stability theory, the complex-valued $\omega(k)=\omega_r(k) + i\omega_i(k)$ is examined, where the function $\omega_i(k)$ determines the exponential growth in time of a mode. The maximum growth rate over the range of $k\in (-\infty,\infty )$, denoted as $\omega_{i,\textrm{max}}$, is used to characterize the stability of the flow. At large times, the exponential nature of the responses~(\ref{eq:mode}) dictate that growth rates at other wavenumbers are subdominant, and the system thus grows as

\begin{subequations}
\begin{equation}
h\sim Ce^{\omega_{i,\textrm{max}}t},\textrm{ as }t\to\infty,
\label{eq:3a}
\end{equation}
where $C$ is a constant. Thus, once $\omega = \omega(k)$ is established from~(\ref{eq:generic}), the linear (exponential) stability is determined as follows~\cite{Chandrasekhar,HuerreRossi,huerre2000}:
\begin{equation}
\omega_{i,\textrm{max}}<0\textrm{: the flow is linearly stable}
\label{eq:3b}
\end{equation}
\begin{equation}
\omega_{i,\textrm{max}}>0\textrm{: the flow is linearly unstable}
\label{eq:3c}
\end{equation}
\begin{equation}
\omega_{i,\textrm{max}}=0\textrm{: the flow is neutrally stable}.
\label{eq:3d}
\end{equation}
\label{eq:classical}
\end{subequations}
The classification~(\ref{eq:classical}) was used by~\cite{Rayleigh1880}, and this classical stability theory has been further developed over the last 100+ years.

The focus of this paper is the classification of neutral stability according to~(\ref{eq:3d}), and is motivated by~\cite{king2016}, henceforth referred to as KRK (the acronym for its first author).  In that work, the classical stability of a fluid flow system yields modes that are neutrally stable for all values of $k$, i.e., $\omega_{i,\textrm{max}}$ = 0 for all wave numbers.  While the classical stability assessment~(\ref{eq:3d}) indicates there will be no growth or decay in a system response, KRK demonstrates both numerically and analytically that disturbances can grow algebraically.  Algebraic growth is defined as a system response that obeys $h\sim Ct^s$, where $C$ is a constant and the exponent $s$ is a positive rational number.  It should be noted that algebraic decay of disturbances ($s<0$) has also been identified in previous work when exponential modes are neutrally stable for all real $k$, having both integer~\cite{case1960} and non-integer~\cite{whitham2011,lighthill2001,barlow2010} character.  It is apparent that the classification of neutral stability via classical means is deficient and warrants further study.  

As is evident in the work of KRK and many earlier studies~\cite{deluca1997,barlow2011} (see KRK for a comprehensive literature review), algebraic growth in linear PDE operators may be examined via a spatio-temporal formulation involving both Fourier and Laplace transforms.  When the inverse Laplace transform is taken first, the resulting Fourier inversion integral has an integrand with an exponential term identical to~(\ref{eq:mode}) with $\omega=\omega(k)$ according to the dispersion relation~(\ref{eq:generic}).  As such, the connection between classical stability analysis and a spatio-temporal analysis is made — the assumption being that the growth characteristics of the superposition of modes of the form~(\ref{eq:mode}) invoked via integration mimic exactly that of the individual modes~(\ref{eq:mode}) when taken separately.  In fact, KRK shows that this is not always the case.  When the exponent $s$ in $h\sim Ct^s$ is fractional, algebraic growth can only be deduced via a superposition of modes via integration, and is thus in fact “non-modal”.  An additional important feature of systems exhibiting algebraic growth is their sensitivity to the type of perturbations to the system.  KRK shows that if perturbations in initial velocity and forcing are invoked, a system response may grow algebraically; however, if the same system is perturbed in location, the system response can decay algebraically.  The sensitivity to initial conditions makes it imperative to consider a broad set of perturbations to a system when establishing system stability, especially so when examining the possibility of algebraic growth.

\begin{figure*}
\begin{center}
\subfloat{(a)\includegraphics[width=2.75in]{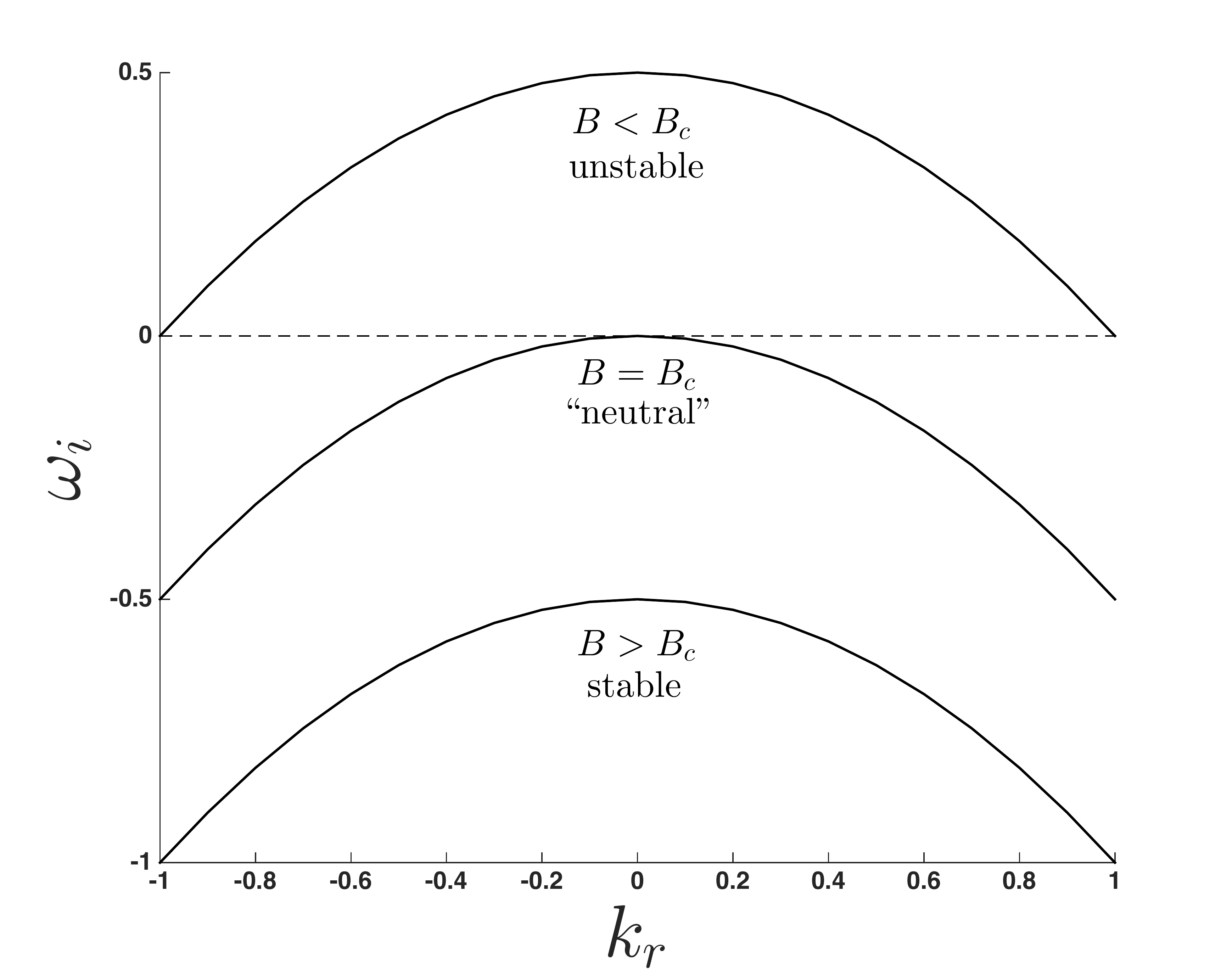}}\\
\subfloat{(b)\includegraphics[width=2.75in]{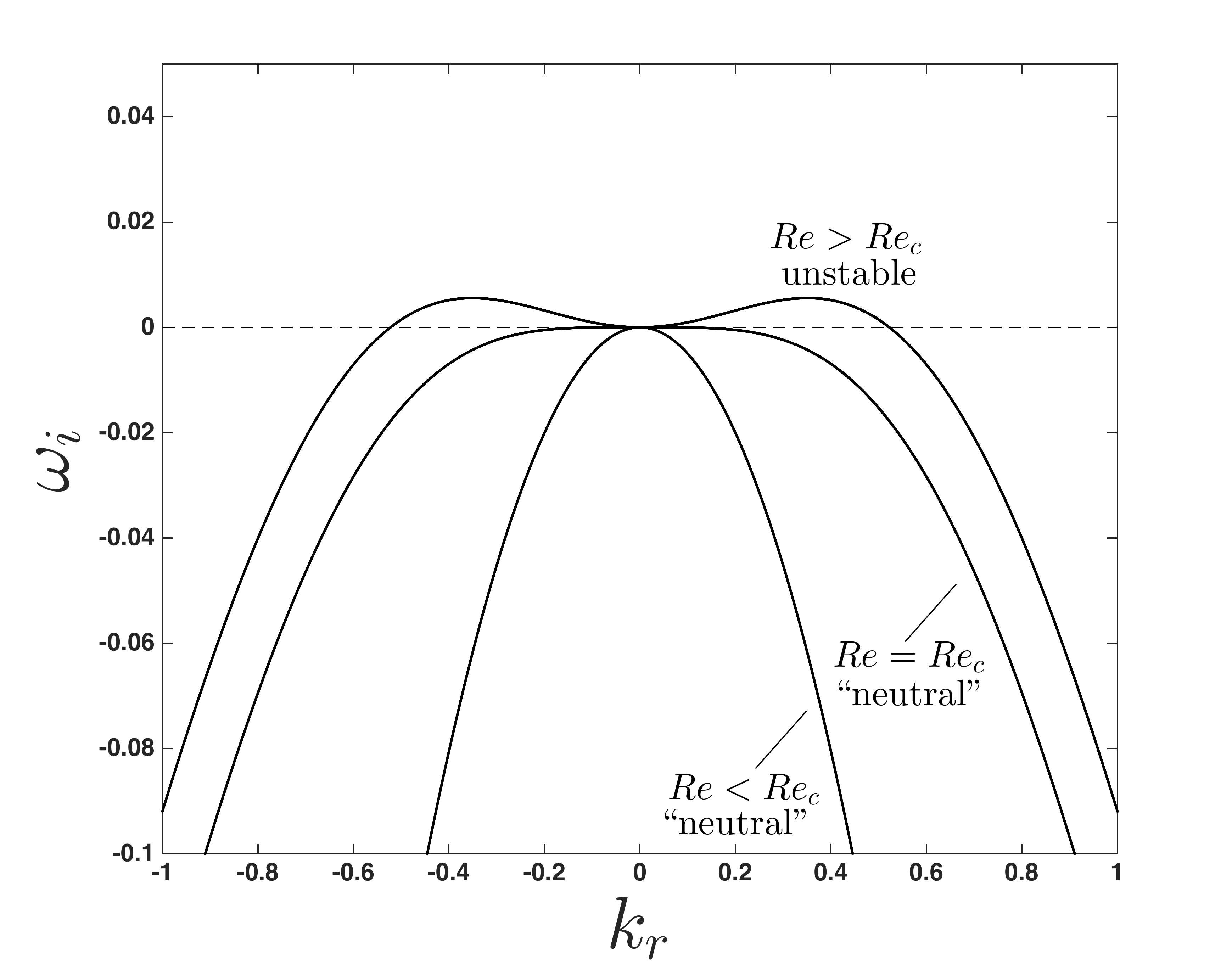}}
\end{center}
\caption{Classical stability curves for two illustrative problems where transition occurs due to the variation of a critical parameter: (a) The transition from exponential growth to exponential decay for the model problem in section~\ref{sec:growth} as the parameter $B$ is varied. The curves (top to bottom) in correspond to $B$=$-1$, 0, and 1 (b) The effect of Reynolds number (defined in section~\ref{sec:decay}) on the stability transition for flow down an inclined plane. The curves (top to bottom) correspond to $R\!e$=1.1 $\cot\theta$, $\cot\theta$, and 0.5 $\cot\theta$ where (for all curves) the incline angle is $\theta=\pi/4$ and Weber number (defined in section~\ref{sec:growth}) is 0.1. }
\label{fig:transition}
\end{figure*}

KRK focuses on a breakdown in classical stability theory for a dispersion relation~(\ref{eq:generic}) in which all modes exhibit neutral stability, yet growth or decay is actually predicted.  The more typical situation in which neutral stability arises in the literature is where a single mode is neutrally stable, and modes corresponding to all other real wavenumbers exhibit damping according to the form~(\ref{eq:mode}) with dispersion relation~(\ref{eq:generic}).  It is widely held in prior literature~\cite{Chandrasekhar,Schlichting,drazin2004,schmid2001} that flows exist in a state that is neither stable nor unstable at a neutral stability boundary from classical theory. A question arises as to whether the breakdown in classical stability theory~(\ref{eq:classical}) reported by KRK extends to the situations shown in Fig.~\ref{fig:transition} when only one wavenumber mode is neutrally stable.  

Fig.~\ref{fig:transition}a is a schematic showing a scenario where this neutral stability configuration occurs (see, for example~\cite{Manneville,Joulin}); here, as a parameter (here, $B$) is varied through its critical value, the flow changes from stable, to neutral, to unstable in accordance with the characterization~(\ref{eq:classical}). The features of such a transition is considered in section~\ref{sec:growth} of this paper.  Fig.~\ref{fig:transition}b shows another type of transition that occurs in the well-studied stability of a single layer Newtonian fluid flowing down an inclined plane, where the critical parameter is the Reynolds number, $R\!e$, defined in section~\ref{sec:decay}.  As indicated in Fig.~\ref{fig:transition}b, unstable flows for $R\!e>R\!e_c$ have a maximum modal growth rate at finite wave number, and as $R\!e\to R\!e_c$ this maximum growth rate diminishes to zero and its associated wavenumber simultaneous approaches $k=0$.  For $R\!e\le R\!e_c$, the $k=0$ mode is neutrally stable, and all other modes are damped — in this case, there is not a true transition from instability to stability based on a classical stability analysis, but rather a transition from instability to neutral stability.  This latter assessment has not been explored in prior literature, as it is widely accepted that when $R\!e<R\!e_c$, inclined plane flow is stable~\cite{Yih,brevdo}.  It is possible that the $k=0$ wavenumber is neglected in prior stability work because it coincides with an interfacial perturbation that is flat, and thus may be viewed as degenerate.  The objective of this paper is to consider the prospect of algebraic growth and decay for the classically neutral stability configurations shown in Figs.~\ref{fig:transition}a and~\ref{fig:transition}b (at $B = B_c$ and $R\!e=R\!e_c$, respectively). 

The paper is organized as follows.  Section 2 introduces a model PDE for the transition shown in Fig.~\ref{fig:transition}a, where it is shown that algebraic growth can occur on the threshold of neutral stability.  The algebraic growth is extracted via long-time asymptotic analysis of the Fourier integral solution, which is then compared with the Fourier series solution of the PDE.  The spatio-temporal classification of algebraic growth as being absolutely or convectively unstable is examined.  In section 3, the same approach is used to examine the well-known incline plane flow studied by~\cite{Yih}.  Here algebraic decay is shown to occur on the neutral stability threshold.  In section 3.1, the governing PDE is derived in non-dimensional form.  In section 3.2, the classical stability analysis is reviewed and key elements are extracted regarding the neutral stability threshold. In section 3.3, the integral solution is examined via asymptotic analysis and the algebraic decay rate is deduced.  A summary of our results and concluding remarks are given in section 4. Key supporting analyses are provided in Appendices A-C.

\section{Model Problem that exhibits algebraic growth~\label{sec:growth}}
In this section, we focus on waves with vertical displacement $h(x,t)$, described by the following PDE, 
\begin{eqnarray}
	&\frac{\partial^2 h}{\partial t^2}+
	2c\frac{\partial^2 h}{\partial x \partial t}+ 
	c^2\frac{\partial^2 h}{\partial x^2}+
	\frac{\partial^4 h}{\partial x^4}-
	\frac{\partial^3 h}{\partial x^2 \partial t}-
	c\frac{\partial^3 h}{\partial x^3}+
	Bc\frac{\partial h}{\partial x}+
	B\frac{\partial h}{\partial t} + B^2h = f_0\delta(x)\delta(t) \nonumber\\
	&h(x,0)=h_0\delta(x),~~\frac{\partial h}{\partial t}(x,0)=u_0\delta(x),~~-\infty<x<\infty,~~t>0,
	\label{eq:growthPDE}
\end{eqnarray}
where $x$ is the horizontal coordinate, $t$ is time, $B$ is an instability parameter and $c$ is a convective parameter.  In~(\ref{eq:growthPDE}), the parameters $h_0$, $u_0$, and $f_0$ are magnitudes of the initial conditions and forcing.  The use of the delta function as an efficient means to initiate disturbances is well-established~\cite{king2016}.  The PDE~(\ref{eq:growthPDE}) was not physically motivated; rather, it was reverse engineered from a dispersion relation that leads to the ``textbook''~\cite{Manneville} depiction of instability transition as shown in Fig.~\ref{fig:transition}a.  This led to the expression~(\ref{eq:growthPDE}) but with $c=0$.  The parameter $c$ was later added to incorporate an element of convection without affecting the stability character of Fig.~\ref{fig:transition}a. 

\subsection{Classical stability analysis \& comparison with Fourier series}
\begin{figure*}
\begin{center}
\subfloat{(a)\includegraphics[width=3.3in]{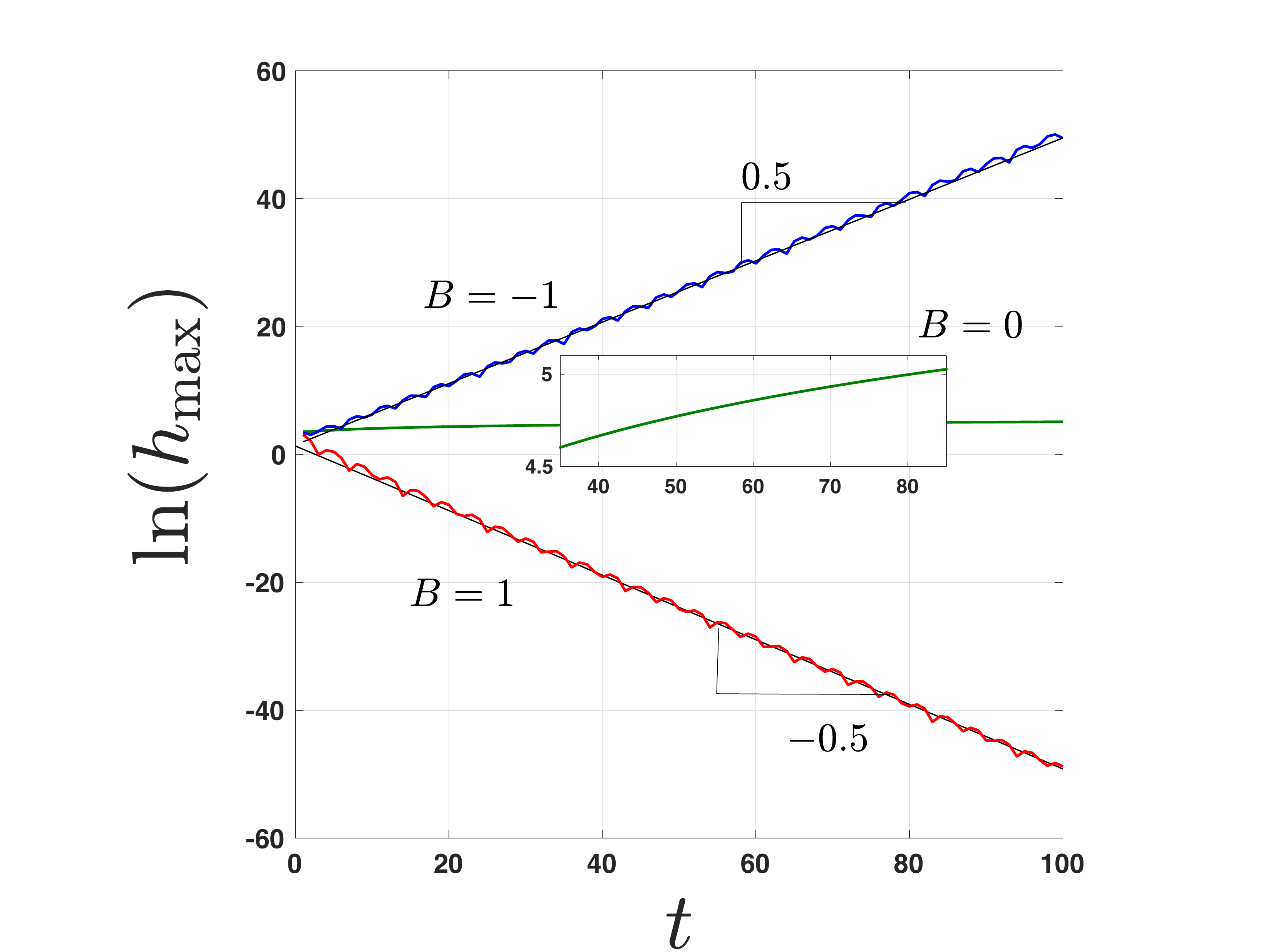}}\\
\subfloat{(b)\includegraphics[width=3.3in]{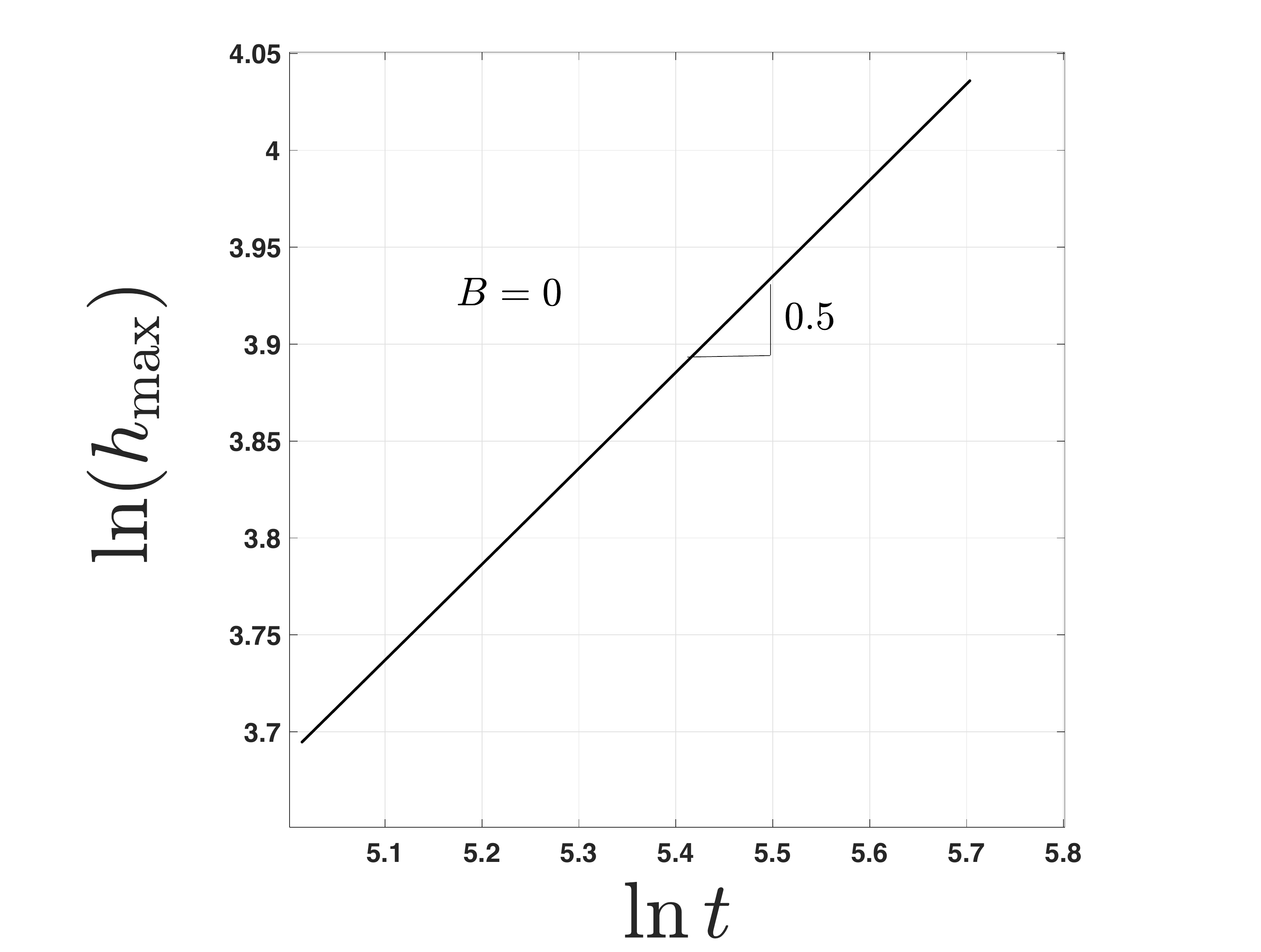}}
\end{center}
\caption{(a) The evolution of the magnitude of the wave peak vs. time in the Fourier series solution of~(\ref{eq:growthPDE}) for $c=0$ and various $B$ values. The slopes on this semi-log plot confirm the exponential growth rates predicted by classical stability analysis. (b) The $B=0$ case from (a) instead shown on a log-log plot, indicating algebraic growth as $t^{1/2}$. }
\label{fig:logplots}
\end{figure*}

Substituting $h=C{{e}^{i(kx-\omega t)}}$ into the homogeneous version of~(\ref{eq:growthPDE}) leads to the dispersion relation
\begin{equation}
\omega = \frac{-2Ck+(k^2+B)i \pm \sqrt{3k^4-2k^2B +3B^2}}{-2},
\label{eq:dispersion1}
\end{equation}
whose imaginary part $\omega_i$ is plotted versus real $k$ in Fig.~\ref{fig:transition} for $C=0$ and $B$=$-1$ (top curve), 0 (middle), and 1 (bottom), indicating (classical) temporal growth rates of $\omega_{i,\textrm{max}}$=0.5, 0, and $-0.5$ respectively.  These (exponential) growth rates are validated by the Fourier series solution of~(\ref{eq:growthPDE}), as shown in Fig.~\ref{fig:logplots}a where the peak of the impulse response is plotted versus time.   It is clear from the figure that the classical predictions of exponential growth ($B=-1$,  $\omega_{i,\textrm{max}}$=0.5) and decay  ($B=1$,  $\omega_{i,\textrm{max}}=-$0.5) accurately describe the long-time behavior of the response.  However, for the classically neutral case of $B=0$, an exponential growth rate of $\omega_{i,\textrm{max}}$=0 does not, in fact, yield zero growth, as shown by the inset in Fig.~\ref{fig:logplots}a.   The log-log plot of Fig.~\ref{fig:logplots}b indicates that the peak grows like $\approx t^{1/2}$ when $B=0$.  We now proceed to show that the algebraic growth rate is indeed exactly 1/2 via asymptotic analysis of the integral solution of~(\ref{eq:growthPDE}).

\subsection{Exact and asymptotic solutions\label{sec:GrowthExact}}

Here we focus solely the case of classical neutral stability as shown in Fig.~\ref{fig:logplots}  where $B=B_c=0$.  In this case, the system ~(\ref{eq:growthPDE}) becomes:
\begin{eqnarray}
	\frac{\partial^2 h}{\partial t^2}+
	2c\frac{\partial^2 h}{\partial x \partial t}+ 
	c^2\frac{\partial^2 h}{\partial x^2}+
	\frac{\partial^4 h}{\partial x^4}-
	\frac{\partial^3 h}{\partial x^2 \partial t}-
	c\frac{\partial^3 h}{\partial x^3}=f_0\delta(x)\delta(t),
	\nonumber\\
	h(x,0)=h_0\delta(x),~~\frac{\partial h}{\partial t}(x,0)=u_0\delta(x),~~-\infty<x<\infty,~~t>0,
	\label{eq:neutralPDE}
\end{eqnarray}
To solve the system~(\ref{eq:neutralPDE}), the Fourier transform and its inverse are utilized; these are given respectively by equations~(\ref{eq:18a}) and~(\ref{eq:18b}) as follows
\begin{equation}
\hat{h}\left( k,t \right)=\int\limits_{-\infty }^{\infty }{h(x,t){{e}^{-ikx}}dx},
\label{eq:18a}
\end{equation}
\begin{equation}
h\left( x,t \right)=\frac{1}{2\pi}\int\limits_{-\infty }^{\infty }{\hat{h}(k,t){{e}^{ikx}}dx}.
\label{eq:18b}
\end{equation}
The Fourier transform of~(\ref{eq:neutralPDE}) yields
\begin{eqnarray}
\frac{d^2 \hat{h}}{d t^2}+\left(k^2+2ick\right)\frac{d \hat{h}}{d t}+\left(k^4+ick^3-c^2k^2\right)\hat{h}=f_0\delta(t),
\nonumber\\
	\hat{h}(0)=h_0,~~\frac{d \hat{h}}{d t}(0)=u_0,~~t>0,
	\label{eq:FT}
\end{eqnarray}
where $\hat{h}(k,t)$ denotes the Fourier transform of $h(x,t)$.  Equation~(\ref{eq:FT}) is a linear constant coefficient ordinary differential equation,a whose solution is
\begin{eqnarray}
&\hat{h}(k,t)=e^{-ickt}\left[F_1(k)e^{-r_1k^2t}+F_2(k)e^{-r_2k^2t}\right],\nonumber\\
&F_{1,2}(k)=\frac{\mp (u_0+f_0)i}{\sqrt{3}k^2}\pm\frac{h_0c}{\sqrt{3}k}\mp\frac{h_0}{6}\left(i\sqrt{3}\mp3\right),~~r_{1,2}=\frac{1\mp i\sqrt{3}}{2}
\label{eq:ODEsoln}
\end{eqnarray}
The inverse Fourier transform of~(\ref{eq:ODEsoln}) is
\begin{eqnarray}
\displaystyle
h(x,t)&=\frac{1}{2\pi}\int_{-\infty}^{\infty} \hat{h}(k,t)e^{ikx}~dk,\nonumber\\
&=\frac{1}{2\pi}\int_{-\infty}^{\infty} \left[F_1(k)e^{-r_1k^2t}+F_2(k)e^{-r_2k^2t}\right]e^{ik\left(\frac{x}{t}-c\right)t}~dk,
\label{eq:IFT}
\end{eqnarray}
where the quantity $t$ is factored out of the exponential and the ``velocity'' quantity $x/t$ is introduced to highlight that the structure of~(\ref{eq:IFT}) and associated integration technique to follow depends on whether $x/t=c$ or $x/t\neq c$.  Note that we obtain the same result given by~(\ref{eq:IFT}) if we utilize the Fourier-Laplace double inversion, as done in~\cite{king2016} (and references therein). As is standard for methods such as steepest descent and stationary phase, we construct a long-time asymptotic solution of~(\ref{eq:IFT}) by examining the integrals along a fixed velocities $x/t$ as $t\to\infty$.   Before doing this, we first split~(\ref{eq:IFT}) into 3 integrals, sorted by order of the $k$ singularities in the integrand: 
\begin{eqnarray}
\displaystyle
h(x,t)&=\frac{i(u_0+f_0)}{2\sqrt{3}\pi}\int_{-\infty}^{\infty}\frac{e^{-r_2k^2t}-e^{-r_1k^2t}}{k^2}e^{ik\left(\frac{x}{t}-c\right)t}~dk\nonumber\\
&+\frac{h_0c}{2\sqrt{3}\pi}\int_{-\infty}^{\infty}\frac{e^{-r_1k^2t}-e^{-r_2k^2t}}{k}e^{ik\left(\frac{x}{t}-c\right)t}~dk\nonumber\\
&+\frac{h_0}{12\pi}\int_{-\infty}^{\infty}\left[\left(3-i\sqrt{3}\right)e^{-r_1k^2t}+\left(3+i\sqrt{3}\right)e^{-r_2k^2t}\right]e^{ik\left(\frac{x}{t}-c\right)t}~dk.
\label{eq:three}
\end{eqnarray}
As evidenced by the coefficient of the first integral in~(\ref{eq:three}), the delta function forcing utilized yields a response equivalent to that for  a perturbation in velocity at $t=0$.  It is useful to decompose the imaginary exponentials of the first and third integrals of~(\ref{eq:three}) into sines and cosines via Euler's relation while leaving the second integral as is, since identities for these integrals are readily available~\cite{GR}.  After simplifying the first and third integrals,~(\ref{eq:three}) is rewritten as
\begin{eqnarray}
\displaystyle
h(x,t)=&\frac{2(u_0+f_0)}{\sqrt{3}\pi}\int_{0}^{\infty}\frac{e^{-\frac{1}{2}k^2t}}{k^2}\sin\left(\frac{\sqrt{3}}{2}k^2t\right)\cos\left[k\left(\frac{x}{t}-c\right)t\right]~dk\nonumber\\
&+\frac{h_0c}{2\sqrt{3}\pi}\int_{-\infty}^{\infty}\frac{e^{-r_1k^2t}-e^{-r_2k^2t}}{k}e^{ik\left(\frac{x}{t}-c\right)t}~dk\nonumber\\
&+\frac{h_0}{3\pi}\int_{0}^{\infty}e^{-\frac{1}{2}k^2t}\left[3\cos\left(\frac{\sqrt{3}}{2}k^2t\right)+\sqrt{3}\sin\left(\frac{\sqrt{3}}{2}k^2t\right)\right]\cos\left[k\left(\frac{x}{t}-c\right)t\right]~dk. \nonumber\\
\label{eq:trig}
\end{eqnarray}
The three integrals in~(\ref{eq:trig}) may be evaluated exactly using the identities given by~(\ref{eq:Igeneral}),~(\ref{eq:IntF}), and~(\ref{eq:GRidentities}) provided in Appendix~\ref{sec:identities} to yield
\begin{eqnarray}
\nonumber
h(x,t)=&~\frac{\left(u_0+f_0\right)i\left(\frac{x}{t}-c\right)t}{4\sqrt{3\pi}}\left\{\Gamma\left[-\frac{1}{2},\left(1+\sqrt{3}i\right)\frac{\left(\frac{x}{t}-c\right)^2t}{8}\right]-\Gamma\left[-\frac{1}{2},\left(1-\sqrt{3}i\right)\frac{\left(\frac{x}{t}-c\right)^2t}{8}\right]\right\}\\~&+~\frac{h_0c}{\sqrt{3}}~\textrm{Im}\left[\textrm{erf}\frac{\left(\frac{x}{t}-c\right)\sqrt{t}\left(\sqrt{3}-i\right)}{4}\right]+~\frac{h_0\cos\left[\frac{\left(\frac{x}{t}-c\right)^2t\sqrt{3}}{8}\right]}{\exp\left[\frac{\left(\frac{x}{t}-c\right)^2t}{8}\right]\sqrt{3\pi t}},
\label{eq:exact}
\end{eqnarray}
where $\Gamma$ is the upper incomplete gamma function.

The response~(\ref{eq:exact}) drastically simplifies along the ray $x/t=c$. Applying the limit as $x/t\to c$ to the aggregate first term of~(\ref{eq:exact}) leads to $(u_0+f_0)\sqrt{t/(3\pi)}$. The second term in~(\ref{eq:exact}) is zero for $x/t=c$, and one may directly substitute $x/t=c$ into the third term of~(\ref{eq:exact}) to obtain $h_0/\sqrt{3\pi t}$; collecting these results,~(\ref{eq:exact}) reduces to  

\begin{eqnarray}
h(x,t)|_{\frac{x}{t}=c}=\frac{1}{\sqrt{3\pi}}\left[(u_0+f_0)t^{\frac{1}{2}}+h_0t^{-\frac{1}{2}}\right].
\label{eq:vc}
\end{eqnarray}

 The large time behavior of $h$ for $x/t\neq c$ is found by expanding the first two terms of~(\ref{eq:exact}) as $t\to\infty$; these expansions are found in~\cite{Abramowitz} as Eqns.~6.5.32 and~7.1.23, respectively.  After replacing the first and second terms of~(\ref{eq:exact}) with their leading-order behavior as $t\to\infty$ and then factoring out the decaying exponential appearing in all three terms, we obtain
\begin{eqnarray}
\nonumber
h(x,t)|_{\frac{x}{t}\neq c}\sim\frac{e^{-\left(\frac{x}{t}-c\right)^2t/8}}{\sqrt{3\pi t}}&\left\{\left[\frac{4(u_0+f_0)}{\left(\frac{x}{t}-c\right)^2}-\frac{h_0c}{\left(\frac{x}{t}-c\right)}+h_0\right]\cos\left[\frac{\sqrt{3}\left(\frac{x}{t}-c\right)^2t}{8}\right]\right.
\\&~\left.-~\frac{h_0c\sqrt{3}}{\left(\frac{x}{t}-c\right)}\sin\left[\frac{\sqrt{3}\left(\frac{x}{t}-c\right)^2t}{8}\right]\right\},~~t\to\infty.
\label{eq:vnotc2}
\end{eqnarray}

\begin{figure*}
\begin{center}
\subfloat{\includegraphics[width=3.5in]{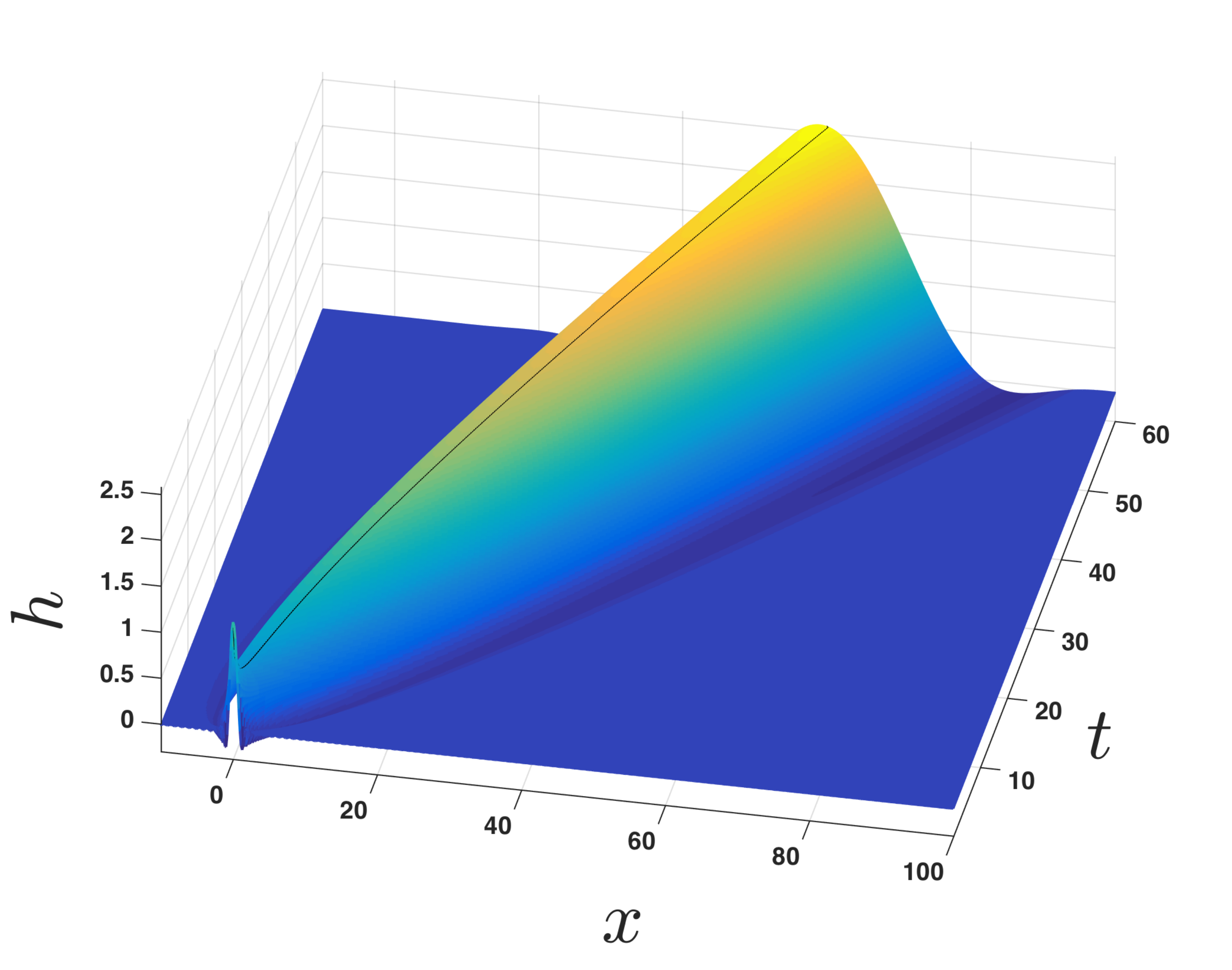}}
\end{center}
\caption{(a) Fourier series solution (shaded surface) to~(\ref{eq:neutralPDE}) compared with the exact solution~(\ref{eq:vc}) (black curve) along the line $x/t=c$.  $h_0=u_0=c=1$, $f_0$=0.  }
\label{fig:growthcompare}
\end{figure*}

 To summarize equations~(\ref{eq:vc}) and~(\ref{eq:vnotc2}), the response grows algebraically like $t^\frac{1}{2}$ for $x/t=c$ (first term of~(\ref{eq:vc}), assuming $u_0+f_0\neq0$) and decays exponentially for $x/t\neq c$ (all terms of~(\ref{eq:vnotc2})). The Fourier series solution to~(\ref{eq:growthPDE}) is shown in Fig.~\ref{fig:growthcompare} for  $u_0+f_0=1$.  All Fourier series solutions displayed in this paper are constructed on a periodic domain, following the approach given in~\cite{barlow2010}.  In Fig.~\ref{fig:growthcompare} the peak is shown to be growing in accordance with the exact solution given by~(\ref{eq:vc}) along $x/t=c$, as indicated by a black line in the figure.

\begin{figure*}
\begin{center}
\subfloat{\includegraphics[width=3.5in]{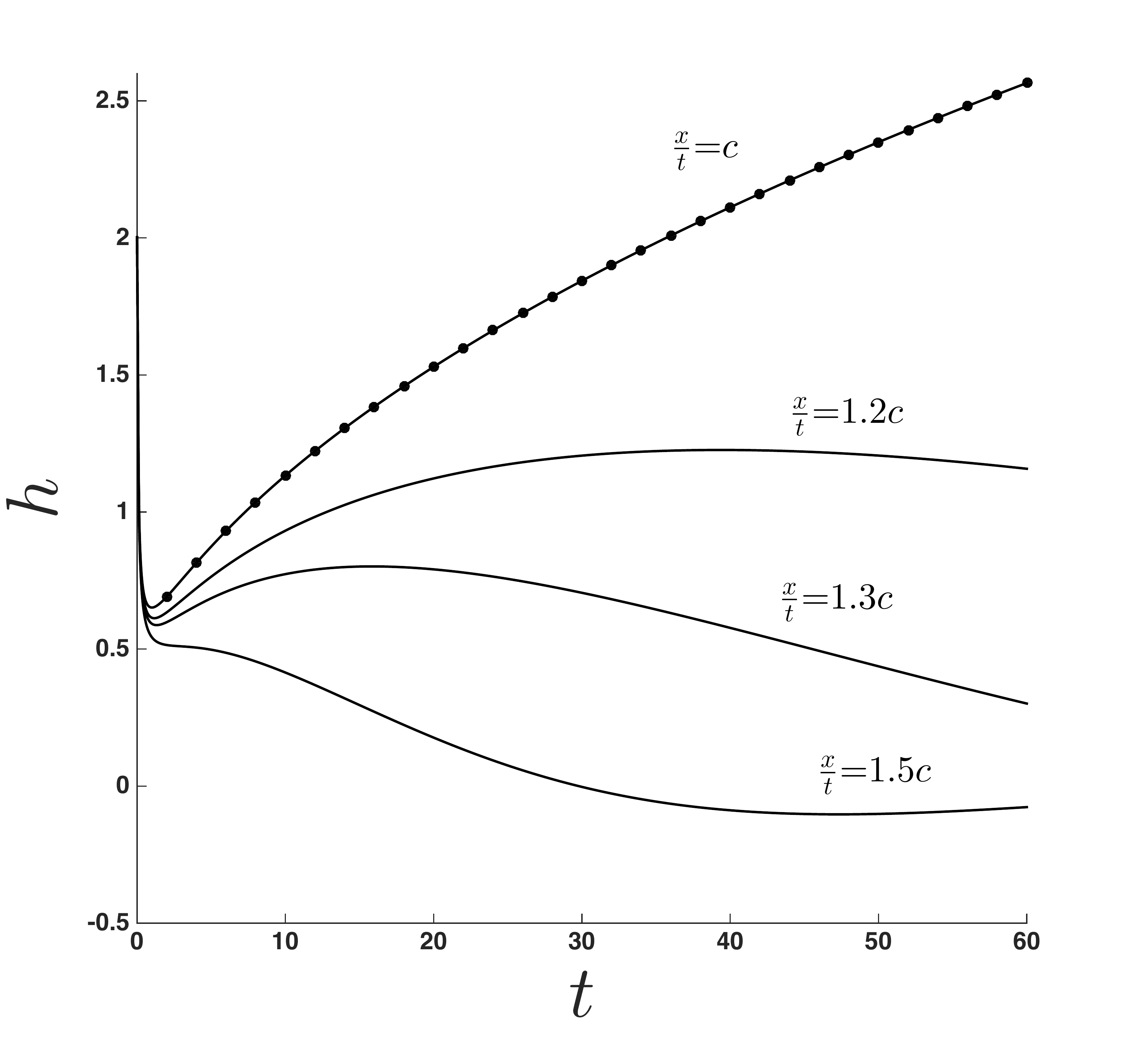}}\\
\subfloat{\includegraphics[width=3.5in]{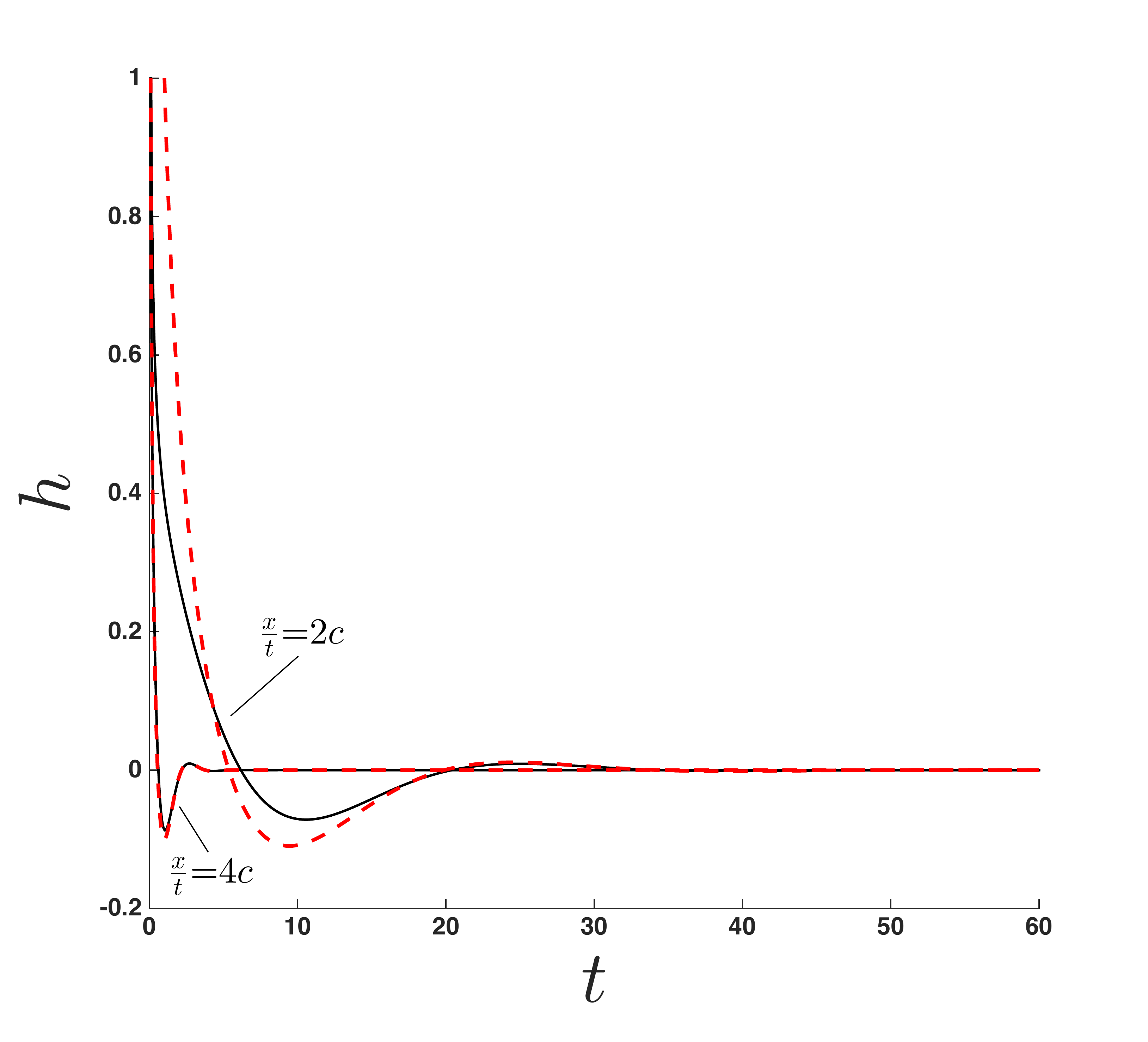}}
\end{center}
\caption{Fourier series solution (solid curves) to~(\ref{eq:neutralPDE}) along specific $x/t$ rays, compared with exact solution given by~(\ref{eq:vc}) (top, denoted by $\bullet$) along $x/t=c$ and leading-order asymptotic solution given by~(\ref{eq:vnotc2}) (bottom dashed curves) for $x/t\neq c$. $h_0=u_0=c=1$, $f_0$=0. }
\label{fig:growthrays}
\end{figure*}

Fig.~\ref{fig:growthcompare} gives the appearance of growth that is spreading as opposed to being confined to a single peak moving at $x/t=c$.  However, the growing response that we see is, in fact, mostly made of transient (``short''-time) growth, which, along any given ray $x/t\neq c$ will eventually damp as $t\to\infty$.  Points moving near the peak of the response at velocities closer to $c$ will take a longer time to damp, as can be seen in Fig.~\ref{fig:growthrays}, where the response is tracked along specific $x/t$ rays and compared with the exact solution ($\bullet$'s) given by~(\ref{eq:vc}) for $x/t=c$ and the long-time asymptotic solution (dashed lines) given by~(\ref{eq:vnotc2}).

Although the growth/decay behavior shown here in Figs.~\ref{fig:growthcompare} and~\ref{fig:growthrays} is similar to the algebraic growth behavior of KRK (where are \textit{all} classical modes are neutral), there is one key difference;  Here, the response along non-growing rays decays exponentially, allowing for the non-oscillatory dome-like structure shown in Fig.~\ref{fig:growthcompare}. For the problem of KRK, such non-growing waves decay \textit{algebraically}, allowing for oscillatory effects to persist, leading to a waveform that has many ripples that spread away from the growing peak (see figures 1-5 in~\cite{king2016}).   Despite these differences, it is worth pointing out that the algebraic growth/decay character in both~(\ref{eq:neutralPDE}) and the problem of KRK arises in the evaluation of Fourier integral solutions that contain removable singularities, such as those shown in~(\ref{eq:trig}).  Such singularities (both here and in KRK) require special attention, as it is often not possible to interpret the integrals as principal values when the integrands are even.  Appendix~\ref{sec:identities} provides key steps to obtain either exact or asymptotic solutions containing algebraic growth/decay.

Note that if $u_0+f_0=0$ and $h_0\neq0$, then the response decays algebraically for $x/t=c$ (second term of~(\ref{eq:vc})) and decays exponentially for $x/t\neq c$. Thus, an impulse can be introduced to a system, either through $u_0$, $f_0$, or $h_0$, and this choice
will affect the algebraic stability or instability of the response.  As shown previously in~\cite{barlow2011} and~\cite{king2016}, algebraically growing systems are particularly sensitive to the choice of initiating disturbance, as further evidenced by the results above. 

Besides being a model problem that illustrates algebraic growth, the PDE given by~(\ref{eq:neutralPDE}) is also useful as a model for either convective instability (if $c\neq0$, see Fig.~\ref{fig:growthcompare}) or absolute instability (if $c=0$), provided that growth is enabled (i.e., $u_0+f_0\neq0$).  Convective instability is defined by a response $h$ that grows and convects for all time (i.e., along a ray $x/t\neq0$) but also decays at any fixed location for large time (i.e., along the ray $x/t=0$)~\cite{HuerreRossi}; this is described, respectively, by~(\ref{eq:vc}) and~(\ref{eq:vnotc2}) for $c\neq0$ and $u_0+f_0\neq0$.  Absolute instability is defined by a response $h$ that grows at any fixed location for large time (i.e., along the ray $x/t=0$)~\cite{HuerreRossi}; this is described, respectively, by~(\ref{eq:vc}) and~(\ref{eq:vnotc2}) for $c=0$ and $u_0+f_0\neq0$.   A comparison between algebraic absolute/convective instability behavior and exponential absolute/convective instability behavior is given in~\cite{king2016}.

\section{Algebraically decaying waves in a thin film along an inclined solid~\label{sec:decay}}

\subsection{Governing equations valid for small interfacial slope}
 
\begin{figure*}
\begin{center}
\includegraphics[width=4.5in]{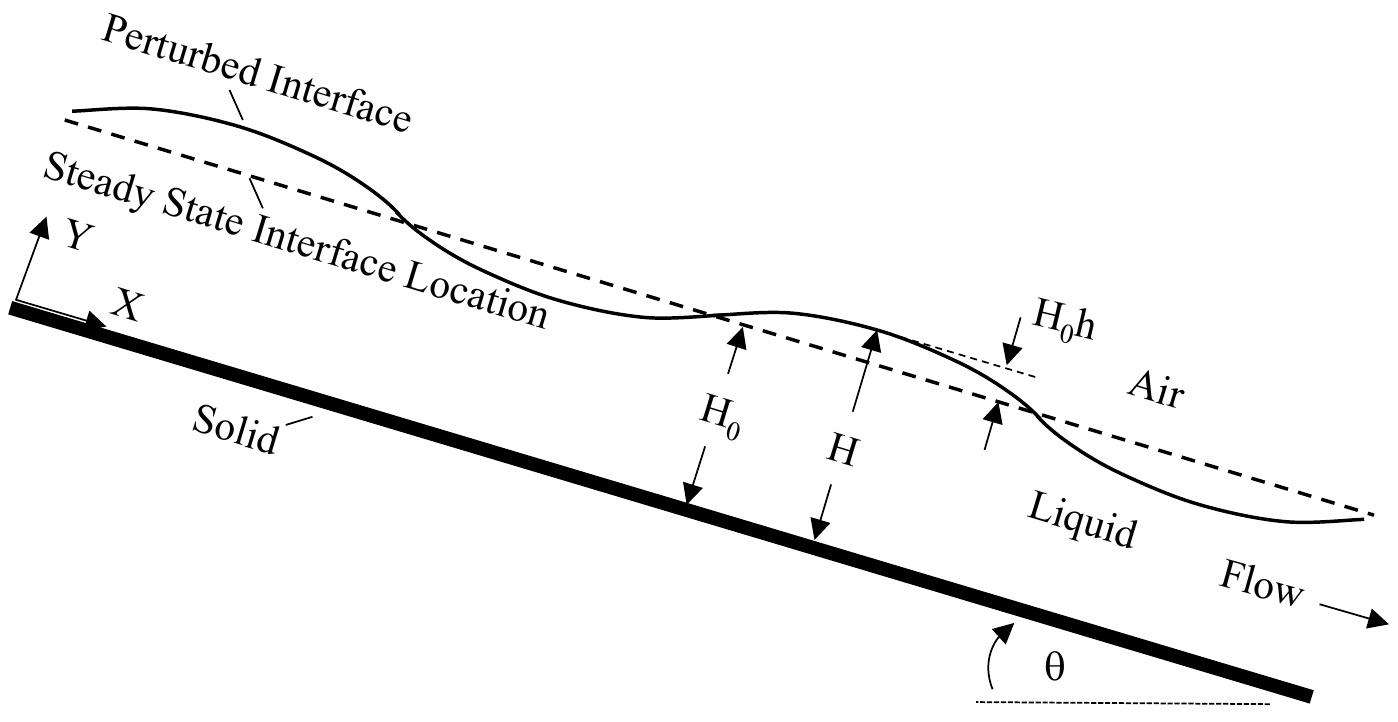}
\end{center}
\caption{Schematic of flow of a thin film along an inclined solid.  The flow is invariant in the $Z$ direction, oriented into and out of the figure.  The steady state interface location is denoted as $H_0$, the perturbed interface location denoted as $H$, and the perturbation magnitude given as $h$, a fraction of $H_0$.}
\label{fig:schematic}
\end{figure*}

We consider a Newtonian liquid of constant density, $\rho$, and constant viscosity, $\mu$, flowing under the influence of gravity, $g$, along a solid surface inclined to horizontal with angle $\theta$ as shown in Fig.~\ref{fig:schematic}.  The liquid layer is exposed to air having a constant atmospheric pressure, and the air-liquid interface has a constant surface tension, $\sigma$.  Under ideal conditions, the liquid flows with a steady-state constant thickness $H_0$ (Fig.~\ref{fig:schematic}) and constant volumetric flow rate per unit width, $Q_0$. The film is perturbed away from uniform due to external disturbances (to be specified later) while remaining invariant in the $Z$ direction (out of Fig.~\ref{fig:schematic}), and the total film thickness deviates from $H_0$ in accordance with the parameterization $Y=H_0h(X,T)$, where $h$ is a dimensionless multiplier, $Y$ is the distance perpendicular to the wall into the fluid domain, $X$ is the distance down the incline, and $T$ is time.  The dynamics of the air are neglected and the external pressure remains atmospheric for all time.  

A set of approximate dynamical equations that govern this configuration may be developed as follows~\cite{Alekseenko,weinstein2004}.  It is assumed that the slope of the perturbed air-liquid interface and underlying fluid trajectories are small.  However, in contrast to lubrication theory that utilizes such assumptions, inertial effects are retained since the flow may be rapid.  The simplified time-dependent Navier-Stokes equations and continuity equation are integrated across the film thickness $H$, and boundary conditions at the wall (no slip and kinematic conditions), and interface (dynamic condition simplified for small slope, kinematic condition) are applied.  The result is an integral equation equivalent to the simplified system of equations and boundary conditions.  Following the approach pioneered by Von Karmen/Polhausen in their treatment of boundary layer theories~\cite{vonKarman,Pohlhausen,Schlichting}, a velocity profile parabolic in $Y$ (see coordinate system in Fig.~\ref{fig:schematic}) is assumed that satisfies the wall boundary conditions, a no shear condition at the unknown interface location $H(X,T)$, and integrates to yield the local volumetric flow per width, $Q(X,T)$.  The resulting system of equations may be written in dimensionless form as~\cite{Alekseenko}:
\begin{subequations}
\label{eq:11}
\begin{equation}
\label{eq:11a}
\frac{\partial \bar{Q}}{\partial x}+\frac{\partial \bar{H}}{\partial t}=0
\end{equation}
\begin{equation}
\label{eq:11b}
\frac{\partial \bar{Q}}{\partial t}+\frac{6}{5}\frac{\partial }{\partial x}\left( \frac{{{{\bar{Q}}}^{2}}}{{\bar{H}}} \right)=\frac{3}{R\!e}\left( \bar{H}-\frac{{\bar{Q}}}{{{{\bar{H}}}^{2}}}-\bar{H}\frac{\partial \bar{H}}{\partial x}\cot \theta  \right)+\frac{1}{W\!e}\bar{H}\frac{{{\partial }^{3}}\bar{H}}{\partial {{x}^{3}}}
\end{equation}
where:
\begin{equation}
\label{eq:11c}
\bar{H}=\frac{H}{{{H}_{0}}},~\bar{Q}=\frac{Q}{{{Q}_{0}}},~x=\frac{X}{{{H}_{0}}},~t=\frac{T{{Q}_{0}}}{{{H}_{0}}^{2}},~R\!e=\frac{\rho {{Q}_{0}}}{\mu },~W\!e=\frac{\rho {{Q}_{0}}^{2}}{\sigma {{H}_{	0}}}.
\end{equation}
Note that $H_0$ and $Q_0$ are not independent and are related as follows
\begin{equation}
H_0=\left(\frac{3Q_0\mu}{\rho g \sin\theta}\right)^{1/3},
\end{equation}
\end{subequations}	
where $H_0$ is the exact expression for the steady-state thickness for film flow calculated from the Navier-Stokes Equations~\cite{Fox}.

The nonlinear system~(\ref{eq:11}) can further be simplified by restricting attention to small interfacial perturbations, which is justified in practical applications where highly uniform films are desired~\cite{weinstein2004}.  The following forms are assumed:
\begin{equation}
\bar{H}\sim1+h(x,t),~\bar{Q}\sim1+q(x,t),~h<<1\textrm{ and } q<<1.
\label{eq:12}
\end{equation}
Equation~(\ref{eq:12}) is substituted into the system~(\ref{eq:11}) and terms quadratic or higher in the perturbation quantities $h$ and $q$ are neglected.  The two linearized equations corresponding to~(\ref{eq:11a}) and~(\ref{eq:11b}) are combined into a single equation to yield:
\begin{subequations}
\label{eq:13}
\begin{equation}
\frac{{{\partial }^{2}}h}{\partial {{t}^{2}}}+\frac{3}{R\!e}\frac{\partial h}{\partial t}+\frac{12}{5}\frac{{{\partial }^{2}}h}{\partial x\partial t}+\frac{1}{W\!e}\frac{{{\partial }^{4}}h}{\partial {{x}^{4}}}+\left( \frac{6}{5}-\frac{3\cot \theta }{R\!e} \right)\frac{{{\partial }^{2}}h}{\partial {{x}^{2}}}+\frac{9}{R\!e}\frac{\partial h}{\partial x}=0.
\label{eq:13a}
\end{equation}
This is the desired governing equation that will be examined in what follows. In accordance with the geometry in Fig.~\ref{fig:schematic}, the spatial domain that will be considered is $-\infty<x<\infty$, and the following boundary conditions and initial conditions are applied:
\begin{equation}
h={{h}_{0}}\delta (x),~\frac{\partial h}{\partial t}={{u}_{0}}\delta (x)\textrm{ at } t=0
\label{eq:13b}
\end{equation}
\begin{equation}
h\to 0\textrm{ as } x\to \pm \infty. 
\label{eq:13c}
\end{equation}
\end{subequations}
In~(\ref{eq:13}), $h_0$ and $u_0$ are constants and $\delta(x)$ is the Dirac delta function.  The system~(\ref{eq:13}) is well posed to solve for the response $h(x,t)$.  Note that, although~(\ref{eq:13a}) is homogeneous, including an impulsive pressure forcing of $f_0\delta(x)\delta(t)$ would have the same response as the initial velocity condition imposed in~(\ref{eq:13b}) as seen for the problem of section 2; thus, we have not incorporated it here.

\subsection{Classical stability analysis}

We now proceed to examine the classical stability of equation~(\ref{eq:13a}).  To do so, boundary conditions~(\ref{eq:13b}) and~(\ref{eq:13c}) are neglected, and the following disturbance form is assumed:
\begin{equation}
h=A{{e}^{i(kx-\omega t)}}.
\label{eq:14}
\end{equation}
In~(\ref{eq:14}), $k$ is a real wavenumber, $\omega=\omega_r+i\omega_i$ is a generally complex frequency, and $A$ is a constant.  Substituting~(\ref{eq:14}) into~(\ref{eq:13a}) and rearranging to assure a non-trivial solution leads to the following:
\begin{subequations}
\begin{equation}
{{\omega }^{2}}+i\beta\omega -\gamma=0.
\label{eq:15a}
\end{equation}
\begin{equation}
\beta=\left( \frac{3}{R\!e}+\frac{12i}{5}k \right),~\gamma=\frac{{{k}^{4}}}{W\!e}-{{k}^{2}}\left( \frac{6}{5}-\frac{3\cot \theta }{R\!e} \right)+\frac{9ik}{R\!e}
\label{eq:15b}
\end{equation}
\label{eq:15}
\end{subequations}
As shown by~\cite{Yih}, the neutral stability condition may be deduced by examining the long wavelength limit as a perturbation series about $k\to0$ holding $c=\omega/k$ fixed, which assures that the speed of any disturbance given by the real part of $c$ is finite.  Equation~(\ref{eq:15}) is first rewritten for fixed $c$ by inserting $\omega=ck$, and the following expansion is inserted into the result:
\begin{equation}
c\sim{{c}_{0}}+{{c}_{1}}k+c_2k^2+c_3k^3+O(k^4)\textrm{ as } k\to 0,~c=\omega/k\textrm{ fixed }
\label{eq:22a}
\end{equation}
where equating like powers leads to
\begin{subequations}
\begin{equation}
c_0=3
\end{equation}
\begin{equation}
{{c}_{1}}=i\left( R\!e-\cot \theta  \right).
\end{equation}
\begin{equation}
{{c}_{2}}=\frac{6}{5} R\!e\left(\cot \theta - R\!e \right).
\end{equation}
\begin{equation}
{{c}_{3}}=i\left[\frac{36}{25}R\!e^2\left(\cot \theta - R\!e \right)-\frac{12}{5}R\!e^3\left(\cot \theta - R\!e \right)^2-\frac{Re}{3W\!e}  \right].
\end{equation}
\label{eq:16}
\end{subequations}
Finally, the results~(\ref{eq:16}) can be rewritten using the definition of $c$ in~(\ref{eq:22a}) to obtain the final form of $\omega$ as:
\begin{eqnarray}
\nonumber
\omega \sim&3k+i(R\!e-\cot \theta )k^2+\frac{6}{5}R\!e\left(\cot \theta - R\!e \right)k^3\\
&+i\left[\frac{36}{25}R\!e^2\left(\cot \theta - R\!e \right)-\frac{12}{5}R\!e^3\left(\cot \theta - R\!e \right)^2-\frac{Re}{3W\!e}  \right]k^4+O(k^5)\textrm{ as }k\to 0.\nonumber\\
\label{eq:17}
\end{eqnarray}
 According to the form~(\ref{eq:14}), the response grows exponentially if $\omega_i>0$, and thus according to~(\ref{eq:17}) for $k\neq0$, there is growth when $R\!e>\cot\theta$.  The approximate nature of equation~(\ref{eq:13}) is revealed here, for when the full linearized Navier-Stokes system is analyzed in the long wavelength limit, the exact result from~\cite{Yih} is that instabilities arise for $R\!e>5/6\cot\theta$; apart from the coefficient difference, the interpretation of instability is identical to that of the approximate analysis.  Note that, to make his stability assessment, Yih only retained terms to order $k^2$ in~(\ref{eq:17}) (i.e., $O$($k$) in~(\ref{eq:22a})).  While the additional terms in~(\ref{eq:17}) do not alter the above stability conclusions, they are required to enable the analysis in section~\ref{sec:FilmAsymptotics} and in supporting analysis found in Appendix~\ref{sec:SD}.
 
\begin{figure*}
\begin{center}
\includegraphics[width=4.5in]{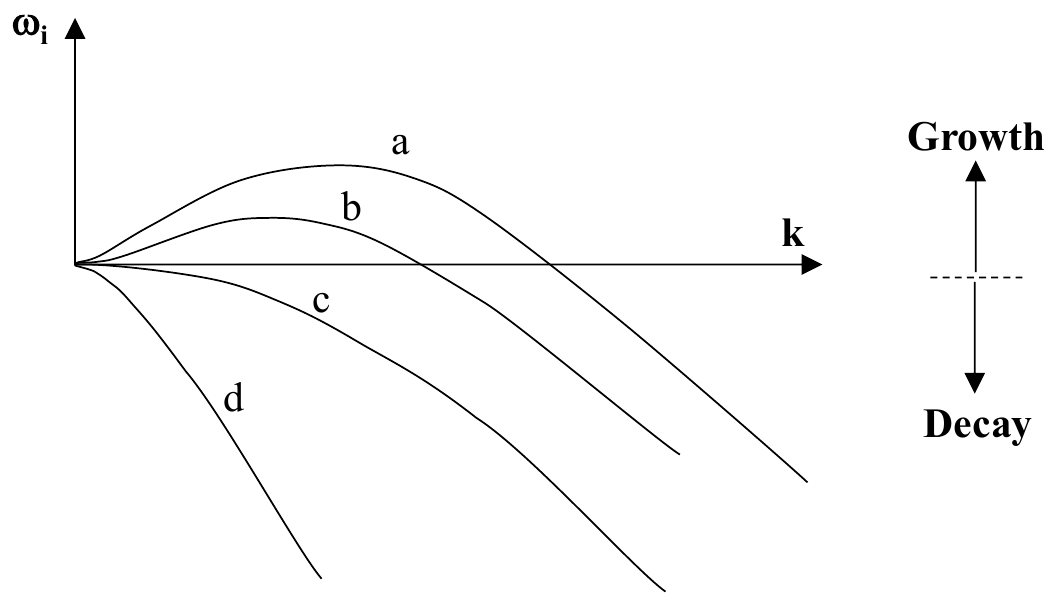}
\end{center}
\caption{Typical plot of growth rate, $\omega_i$, vs real wavenumber, $k$. In the $k\to0$ limit, $\omega_i$ is given by equation~(\ref{eq:17}). Curve a): $R\!e>\cot\theta$, an unstable condition. Note that the maximum growth rate occurs at a finite wavenumber. Curve b): Smaller $R\!e$ than curve a, but still satisfying $R\!e>\cot\theta$. Note that the maximum growth rate is reduced and has shifted to smaller wavenumber. Curve c): The first condition at which there is no growth as $R\!e$ is lowered is $R\!e=\cot\theta$. Curve d): $R\!e<\cot\theta$. For curves c) and d) Modes exponentially damp for all $k\neq0$, and $k=0$ is the wavenumber of maximum (but
zero) growth.}
\label{fig:cartoon}
\end{figure*}

The classical interpretation of exponential growth in the long wavelength limit warrants additional discussion relevant to the current work.  In particular, equation~(\ref{eq:17}) indicates that $\omega_i=0$ when $k=0$, and thus equation~(\ref{eq:14}) shows that this wavenumber exhibits a neutrally stable condition.  Fig.~\ref{fig:cartoon} provides a schematic of the growth rate, $\omega_i$, as a function of $k$, for $k\ge0$, based on solutions of the Orr-Sommerfeld equations (see, for example~\cite{weinstein1990,brevdo}).  As indicated in curve (a), for $R\!e>\cot\theta$, waves exponentially grow over a range of $k$, with the maximum wave growth occurring at finite $k$.  For smaller $R\!e$ that satisfies $R\!e>\cot\theta$ given by curve (b), the wavenumber associated with maximum growth is reduced, until at the neutral condition shown by curve (c), this wavenumber becomes coincident with $k=0$; for all $k\neq0$, modes damp exponentially in accordance with~(\ref{eq:14}). This same structure persists for $R\!e<\cot\theta$.  For all curves shown in Fig.~\ref{fig:cartoon}, note that $k=0$ is in fact a wavenumber that exibits neither growth nor decay, from the classical characterization.  

The Orr-Sommerfeld solutions in Fig.~\ref{fig:cartoon} show why a stability assessment for $R\!e>\cot\theta$ may be made based on the $k\to0$ limit.  Although the maximum growth in this parameter range generally occurs at a finite value of $k$, the concavity of the asymptotic solution~(\ref{eq:17}) is always positive when there is growth; this enables positive concavity in the $k\to0$ limit to assure instability.  For situations where $R\!e=\cot\theta$ or $R\!e<\cot\theta$, the concavity of~(\ref{eq:17}) is zero or negative, respectively.  For these cases, however, the $k=0$ wavenumber precludes a stability conclusion (as pointed out by~\cite{Yih}), since $k=0$ corresponds to the wavenumber of maximum (zero) growth.  We set out to make this assessment in the analysis to follow, where we explicitly examine the system response for cases where $R\!e\le\cot\theta$ via asymptotic analysis.

\subsection{Integral solution and long-time asymptotic behavior \label{sec:FilmAsymptotics}}
We now proceed to solve the system~(\ref{eq:13}) and examine its stability features, following the process discussed in section~\ref{sec:GrowthExact}.  The Fourier transform~(\ref{eq:18a}) of~(\ref{eq:13}) yields 
\begin{subequations}
\begin{equation}
\frac{{{d}^{2}}\hat{h}}{d{{t}^{2}}}+\beta \frac{d\hat{h}}{dt}+\gamma \hat{h}=0.
\label{eq:19a}
\end{equation}
It is implicit in the use of the Fourier transform that equation~(\ref{eq:13c}) is satisfied.  The associated initial conditions~(\ref{eq:13b}) become:
\begin{equation}
\hat{h}={{h}_{0}},~\frac{\partial \hat{h}}{\partial t}={{u}_{0}}\textrm{ at } t=0.
\label{eq:19c}
\end{equation}
\label{eq:19}
\end{subequations}
The ordinary differential system~(\ref{eq:19}) is solved to yield
\begin{equation}
\hat{h}=C_{1}(k)e^{-i\omega_1(k)t}+C_{2}e^{-i\omega_2(k)t},
\label{eq:20}
\end{equation}
where $\omega_1(k)$ and $\omega_2(k)$ are the two roots of the quadratic equation~(\ref{eq:15}) given as
\begin{subequations}
\begin{equation}
{{\omega_1}}(k)=-\frac{i}{2}\left( \beta -\sqrt{\beta^2-4\gamma} \right),~{{\omega_2}}(k)=-\frac{i}{2}\left( \beta +\sqrt{\beta^2-4\gamma} \right)
\label{eq:21a}
\end{equation}
and
\begin{equation}
{{C}_{1}}(k)=\frac{\omega_2h_0-iu_0}{\omega_2-\omega_1},~{{C}_{2}}(k)=\frac{iu_0-\omega_1h_0}{\omega_2-\omega_1}
\label{eq:21c}
\end{equation}
The final solution for $h(x,t)$ may be obtained by utilizing the inverse Fourier Transform~(\ref{eq:18b}) to obtain
\begin{equation}
h\left( x,t \right)=\frac{1}{2\pi}\int\limits_{-\infty }^{\infty }\left[{C}_{1}(k){{e}^{{{\phi }_{1}}(k)t}}+{C}_{2}(k){{e}^{{{\phi }_{2}}(k)t}}\right]dk
\label{eq:21d}
\end{equation}
where
\begin{equation}
{{\phi }_{1}}(k)=i\left[k\frac{x}{t}-{{\omega}_{1}}(k)\right],~{{\phi }_{2}}(k)=i\left[k\frac{x}{t}-{{\omega}_{2}}(k)\right].
\label{eq:21e}
\end{equation}
\label{eq:21}
\end{subequations}
The result~(\ref{eq:21}) provides the integral solution to the system~(\ref{eq:13}), where $\beta$ and $\gamma$ are given in~(\ref{eq:15b}). 

The long time asymptotic behavior of~(\ref{eq:21}) for $R\!e$$\le\cot(\theta)$ may be used to establish the stability of the flow system governed by equation~(\ref{eq:13}).  In Appendix~\ref{sec:balance}, it is shown that the second term in~(\ref{eq:21d}) damps faster than the first as $t\to\infty$ for all $k$, and so we may recast~(\ref{eq:21d})  as 
\begin{equation}
h\left( x,t \right)\sim\frac{1}{2\pi}\int\limits_{-\infty }^{\infty }{C}_{1}(k){{e}^{{{\phi }_{1}}(k)t}}dk,~t\to\infty,~R\!e\le\cot(\theta)
\label{eq:FI}
\end{equation}

Although we cannot determine a closed-form solution for the Fourier integral in~(\ref{eq:FI}), it can be evaluated via the method of steepest descent~\cite{bender1999}, where we allow $k=k_r+ik_i$ to be complex and look for a closed integration path (in the complex $k$-plane) that includes the real line and passes through a saddle point of~(\ref{eq:21e}); this enables the use of Cauchy's theorem.    An $n^\textrm{th}$ order saddle $k_s$  of $\phi_1$ is defined by 
\begin{equation}
\left.\frac{d\phi_1}{dk}\right|_{k_s}=\dots=\left.\frac{d^{n-1}\phi_1}{dk^{n-1}}\right|_{k_s}=0,~\left.\frac{d^{n}\phi_1}{dk^{n}}\right|_{k_s}\neq0,~n>1.
\label{eq:saddle}
\end{equation}
Note that, depending on the direction of approach, saddle points can describe where Real[$\phi_1]$ reaches a maximum \textit{or} minimum.  The deformed integration path of steepest descent moves through the saddle such that the imaginary part of $\phi_1(k)$ is constant and real part of $\phi_1(k)$ attains its \textit{maximum} value at the saddle. This enables an asymptotic expansion about the saddle point, which can be used to deduce the dominant behavior as $t$ becomes large.  The path of ascent, where a minimum of Real[$\phi_1]$ is reached at the saddle, is not useful in the long-time evaluation of Fourier integrals such as~(\ref{eq:FI}).  Note that~(\ref{eq:FI}) contains branch points when $\beta^2=4\gamma$ and so care must be taken such that the aforementioned path does not enclose such points, as this would violate Cauchy's theorem.

Substituting~(\ref{eq:21e}) into~(\ref{eq:saddle}) leads to the relations for $\omega(k)=\omega_r(k)+i\omega_i(k)$
\begin{subequations}
\begin{equation}
\left.\frac{\partial\omega_r}{dk_r}\right|_{k_s}=\frac{x}{t}
\end{equation}
\begin{equation}
\left.\frac{\partial\omega_i}{dk_r}\right|_{k_s}=0.
\end{equation}
\label{eq:wsaddle}
\end{subequations}
Equation~(\ref{eq:wsaddle}) highlights that each saddle point $k_s$ is paired with an $x/t$ ray; it provides a simultaneous set of equations to solve for the real and imaginary parts of $k_s$ for a given $x/t$.  We may deduce the long-time behavior of~(\ref{eq:FI}) along specific $x/t$ values by expanding $\phi_1(k)$ about the corresponding saddles.  Using this expansion within the method of steepest descent outlined above (see Appendix~\ref{sec:SD}), we arrive at the following long-time behavior for~(\ref{eq:FI})
\begin{subequations}
\begin{equation}
h\left( x,t \right)|_{\frac{x}{t}=3}\sim \frac{h_0+u_0~R\!e/3}{2\sqrt{\pi\left[\cot(\theta)-R\!e\right]}}~t^{-\frac{1}{2}},~~R\!e<\cot(\theta)
\label{eq:asymptoticA}
\end{equation}
\begin{equation}
h\left( x,t \right)|_{\frac{x}{t}=3}\sim \frac{1}{4\pi} \Gamma\left(\frac{1}{4}\right)\left[\frac{3~W\!e}{R\!e}\right]^{\frac{1}{4}}\left(h_0+u_0~R\!e/3\right)~t^{-\frac{1}{4}},~~R\!e=\cot(\theta)
\label{eq:asymptoticB}
\end{equation}
\begin{equation}
h\left( x,t \right)|_{\frac{x}{t}\neq3}\sim c_1 t^{-\frac{1}{2}}e^{c_2t}\cos\left(c_3t+c_4\right),~~c_2<0,~R\!e\le\cot(\theta)
\label{eq:offrays}
\end{equation}
\label{eq:asymptotic}
\end{subequations}
where the real-valued parameters $c_1$, $c_2$, $c_3$, and $c_4$ are functions of $x/t$; a more specific form of~(\ref{eq:offrays}) for direct use is given by~(\ref{eq:offrays2})  in Appendix~\ref{sec:SD}.  Note that there is a structural change for $x/t=3$ and $x/t\neq3$ (for all $R\!e$$\le\cot(\theta)$).  For any $R\!e$, $x/t=3$ is the least damped ray, only exhibiting algebraic decay, while all other rays damp exponentially at $t\to\infty$.  There is also a structural change for $R\!e$ $<\cot(\theta)$ and $R\!e$$=\cot(\theta)$ along the ray $x/t=3$.  For $R\!e$$<\cot(\theta)$, the peak of the wave packet (at $x/t=3$) decays like $t^{-1/2}$.   For $R\!e$$=\cot(\theta)$, the peak of the wave packet decays like $t^{-1/4}$.  This is a ramification of the saddle point being 2$^\textrm{nd}$-order for the former and 4$^\textrm{th}$-order for the latter, as shown in Appendix~\ref{sec:SD}; this change of order can also be seen in~(\ref{eq:17}), setting $R\!e$$=\cot(\theta)$.   Here, we have deduced the stability of the system~(\ref{eq:13}) for $R\!e$$\le\cot(\theta)$, in a parameter space where classical stability analysis is inconclusive. 

\begin{figure*}
\begin{center}
\subfloat{\includegraphics[width=4.5in]{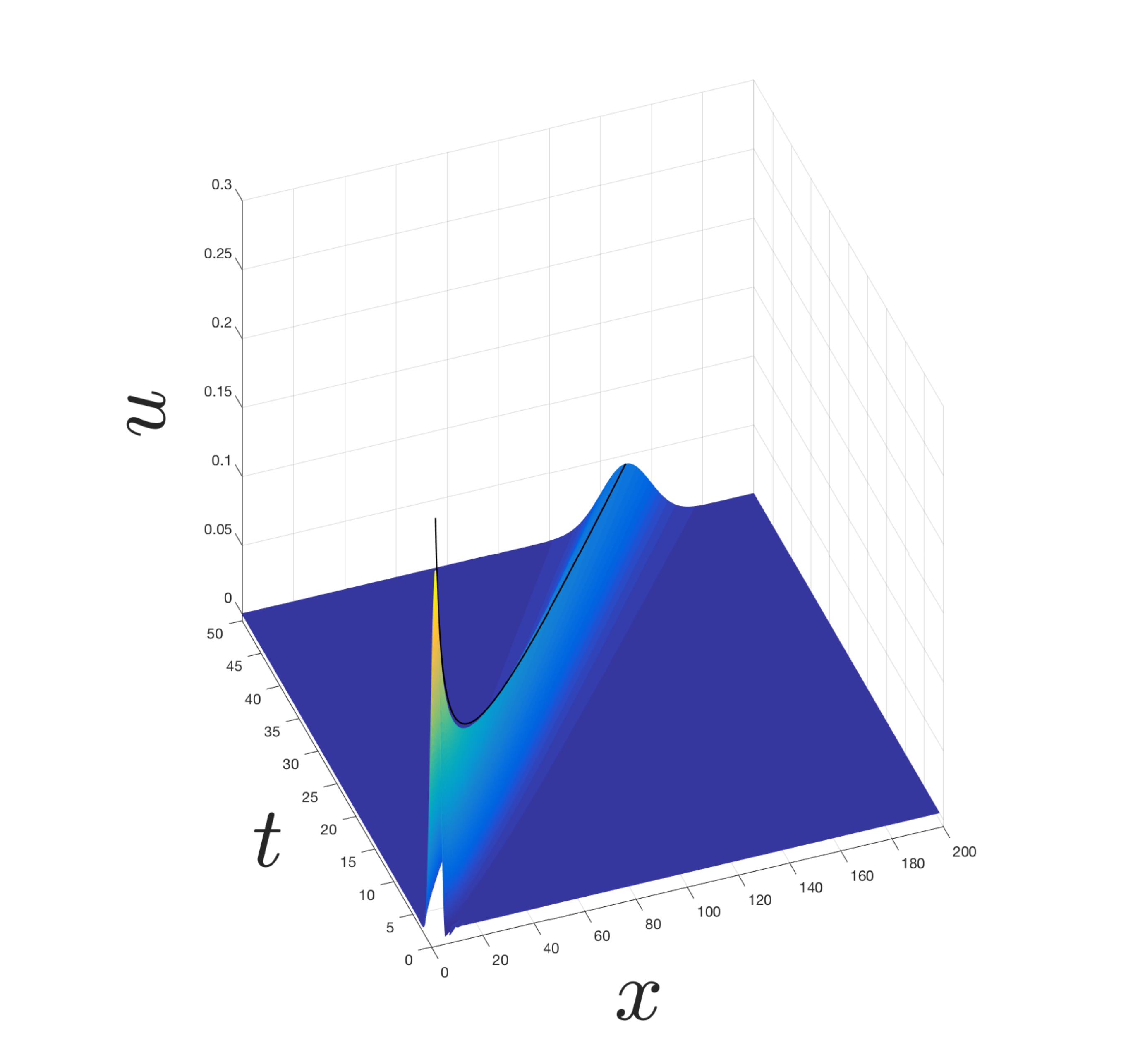}}
\end{center}
\caption{Fourier series solution (shaded surface) to~(\ref{eq:13}) compared with~(\ref{eq:asymptoticB}) (black curve) along the line $x/t=3$. $R\!e$=$\cot(\theta)$, $h_0=u_0=1$, $\theta=\pi/4$, and $W\!e$=0.1.}
\label{fig:meshIncline}
\end{figure*}

\begin{figure*}
\begin{center}
\subfloat{\includegraphics[width=3.9in]{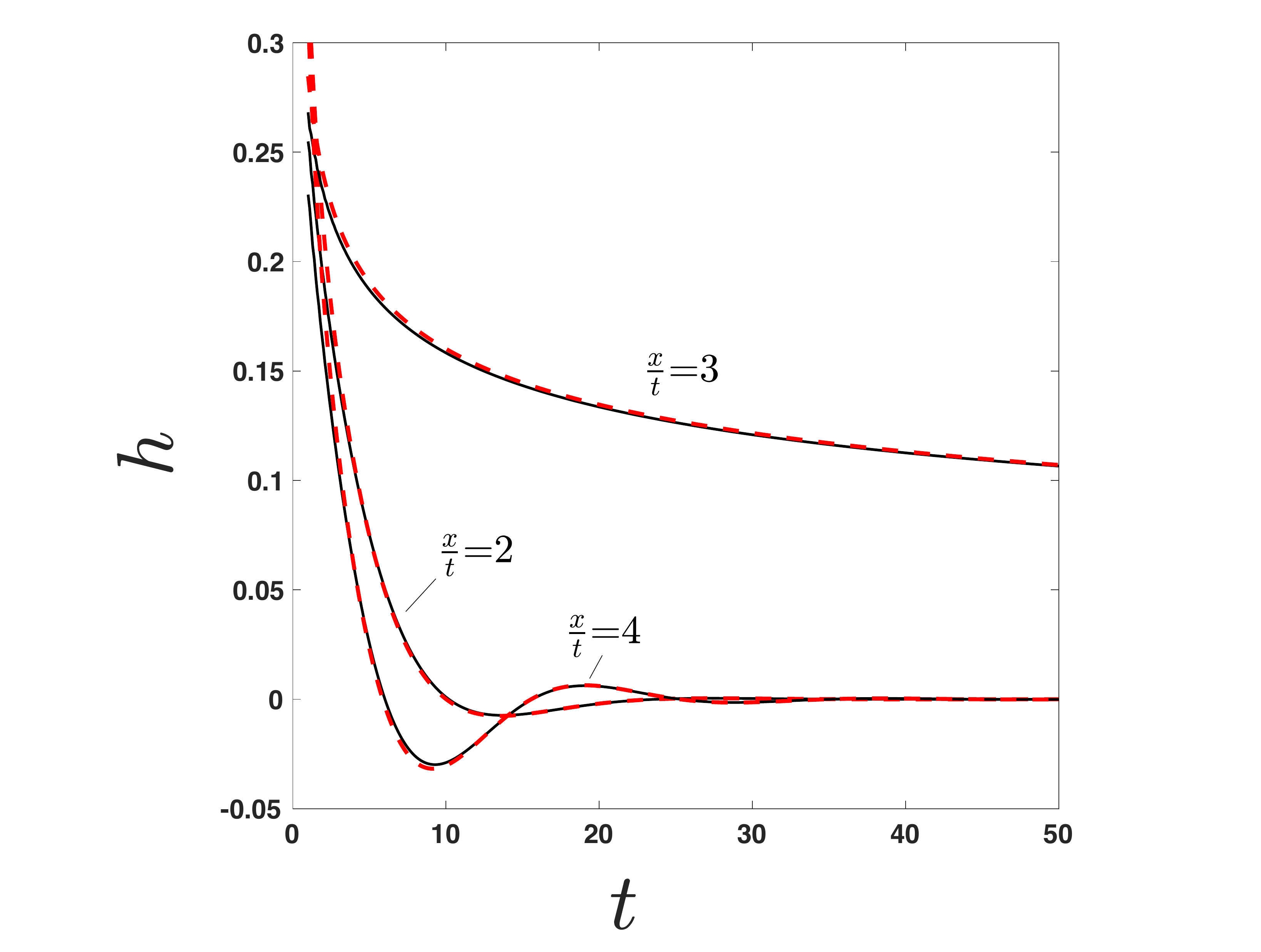}}
\end{center}
\caption{Fourier series solution (solid curves) to~(\ref{eq:13}) along specific $x/t$ rays, compared with the asymptotic solution given by~(\ref{eq:asymptotic}) (dashed curves) for $R\!e$=$\cot(\theta)$, $h_0=u_0=1$, $\theta=\pi/4$, and $W\!e$=0.1.}
\label{fig:decaycompare}
\end{figure*}

The Fourier series solution to~(\ref{eq:13}) is shown in Fig.~\ref{fig:meshIncline}, where the peak is seen to decay in accordance with the asymptotic solution given by~(\ref{eq:asymptoticB}) along $x/t=3$, as indicated by a black line in the figure.  Away from the peak (for $x/t\neq3$) the series solution approaches the asymptotic solution given by~(\ref{eq:offrays}), as indicated in Fig.~\ref{fig:decaycompare}.

\section{Conclusions~\label{sec:conclusions}}
Flow transition from stability to instability is examined in which a single mode (i.e., for one value of wavenumber) indicates classical neutral stability.   It is found that fluid flows characterized as being neutrally stable via exponential modes in fact may exhibit either algebraic growth or decay.  Two partial differential equation systems are examined, one that has been constructed to elucidate key features of the stability threshold, and a second that models the well-studied problem of rectilinear Newtonian flow down an inclined plane.  In the former case, algebraic growth occurs on the neutral stability threshold. In the latter case, algebraic decay occurs both at and below the critical Reynolds number.  

A key difference between the algebraic growth problem presented here and the one in~\cite{king2016} is that here only one wave-number touches the real-axis on a classical stability plot, whereas all wavenumbers are neutral in the prior work.  A ramification of this appears to be that the non-growing rays (away from the peak of the wavepacket) decay exponentially instead of algebraically.  Whether or not this is true for all such problems remains to be explored.   In both problems, algebraic instability arises from removable singularities in the Fourier integral solution of the governing PDE. 

For the second problem reported here \--- that of incline plane flow \--- exponential growth occurs above a critical Reynolds number. In his own work,~\cite{Yih} also reached the conclusion that the flow is unstable above the critical value, but could not reach a conclusion for when the Reynolds number was below the critical value. The discovery of algebraic stability below the critical Reynolds number carries with it the danger that the flow will not be as stable as it would be if all values of $\omega_i$ were negative and disturbances damped exponentially. In the context of coating operations that utilize an inclined plane-flow geometry to form liquid films, our results show that disturbances to a process, even if damped, will damp out more slowly than expected in the physical domain. It is possible that, given larger disturbances, even a damped response may lead to unacceptable perturbations in coated liquid films.

\bibliographystyle{imamat}
\bibliography{neutral}

\appendix
\section{Evaluation of integrals in Eq.~(\ref{eq:trig}) \label{sec:identities}}
\subsection{~\label{sec:gammaint}}
In this section, general formulas for convergent integrals of the following form are established:
\begin{equation}
I=\int_0^\infty\frac{e^{-\beta k^2}}{k^2}\sin(\alpha k^2)\cos\left(bk\right)~dk,~~\beta>0,~~\alpha>0,~~b>0.
\label{eq:I}
\end{equation}
First, we introduce the additional variable $\gamma$ in~(\ref{eq:I}) as follows
\begin{equation}
I=\int_0^\infty\frac{e^{-\gamma\beta k^2}}{k^2}\sin(\gamma\alpha k^2)\cos\left(bk\right)~dk,
\label{eq:I2}
\end{equation}
where $\gamma=1$ in~(\ref{eq:I2}) yields~(\ref{eq:I}).  Next,~(\ref{eq:I2}) is differentiated with respect to $\gamma$ to obtain
\begin{equation}
\frac{dI}{d\gamma}=-\beta\int_0^\infty e^{-\gamma\beta k^2}\sin(\gamma\alpha k^2)\cos\left(bk\right)~dk+\alpha\int_0^\infty e^{-\gamma\beta k^2}\cos(\gamma\alpha k^2)\cos\left(bk\right)~dk.
\label{eq:dI}
\end{equation}
In order to recover the general solution for $I$, we note that~(\ref{eq:I2}) indicates that the following constraint must be satisfied:
\begin{equation}
I=0~\textrm{ at }~\gamma=0.
\label{eq:C}
\end{equation}
The integrals in~(\ref{eq:dI}) are provided in closed-form on pages 493-494, Eqs. (3.922-3) and (3.922-4) of~\cite{GR} (for $\tilde{\beta}>0,~a>0,~b>0$):
\begin{subequations}
\begin{eqnarray}
\int_0^\infty e^{-\tilde{\beta} k^2}\sin(a k^2)\cos\left(bk\right)~dk&=-\sqrt{\frac{\pi}{8\left(\tilde{\beta}^2+a^2\right)}}\textrm{ exp }\left[-\frac{b^2\tilde{\beta}}{4\left(\tilde{\beta}^2+a^2\right)}\right]\nonumber\\
&\times\left\{\sqrt{\sqrt{\tilde{\beta}^2+a^2}+\tilde{\beta}}~\sin\left[\frac{b^2a}{4\left(\tilde{\beta}^2+a^2\right)}\right]\right.\nonumber\\
&\left.-\sqrt{\sqrt{\tilde{\beta}^2+a^2}-\tilde{\beta}}~\cos\left[\frac{b^2a}{4\left(\tilde{\beta}^2+a^2\right)}\right]  \right\}
\label{eq:sine}
\end{eqnarray}

\begin{eqnarray}
\int_0^\infty e^{-\tilde{\beta} k^2}\cos(a k^2)\cos\left(bk\right)~dk&=\sqrt{\frac{\pi}{8\left(\tilde{\beta}^2+a^2\right)}}\textrm{ exp }\left[-\frac{b^2\tilde{\beta}}{4\left(\tilde{\beta}^2+a^2\right)}\right]\nonumber\\
&\times\left\{\sqrt{\sqrt{\tilde{\beta}^2+a^2}+\tilde{\beta}}~\cos\left[\frac{b^2a}{4\left(\tilde{\beta}^2+a^2\right)}\right]\right.\nonumber\\
&\left.+\sqrt{\sqrt{\tilde{\beta}^2+a^2}-\tilde{\beta}}~\sin\left[\frac{b^2a}{4\left(\tilde{\beta}^2+a^2\right)}\right]  \right\}.
\label{eq:cosine}
\end{eqnarray}
\label{eq:GRidentities}
\end{subequations}

Applying~(\ref{eq:sine}) and ~(\ref{eq:cosine}) to~(\ref{eq:dI}), and then integrating~(\ref{eq:dI}) from $\gamma=0$ to 1, we obtain
\begin{eqnarray}
I=\sqrt{\frac{\pi}{8\left(\beta^2+\alpha^2\right)}}\int_0^1\frac{e^{-\beta\theta/\gamma}}{\sqrt{\gamma}}\left[C_+\cos\left(\alpha\theta/\gamma\right)+C_-\sin\left(\alpha\theta/\gamma\right)\right]~d\gamma \nonumber\\
C_\pm=\alpha\sqrt{\sqrt{\beta^2+\alpha^2}\pm\beta}\mp\beta\sqrt{\sqrt{\beta^2+\alpha^2}-\beta},~~\theta=\frac{b^2}{4\left(\beta^2+\alpha^2\right)}.
\label{eq:IntI}
\end{eqnarray}
Upon making the substitution $u=\beta\theta/\gamma$,~(\ref{eq:IntI}) becomes
\begin{equation}
I=\frac{b\sqrt{\beta\pi/2}}{4\left(\beta^2+\alpha^2\right)}\int_{\beta\theta}^\infty u^{-3/2}e^{-u}\left[C_+\cos\left(u\alpha/\beta\right)+C_-\sin\left(u\alpha/\beta\right)\right]~du,
\end{equation}
whose exact solution is given by Eqs. (3.944-2) and (3.944-4) on page 498 of~\cite{GR}, thus providing an exact solution for~(\ref{eq:I}):
\begin{eqnarray}
I=\int_0^\infty&\frac{e^{-\beta k^2}}{k^2}\sin(\alpha k^2)\cos\left(bk\right)~dk=\nonumber\\
&\frac{b\sqrt{\pi/2}}{8\left(\beta^2+\alpha^2\right)}\left\{\left(C_++iC_-\right)\sqrt{\beta+i\alpha}~~\Gamma\left[-\frac{1}{2},\left(\beta+i\alpha\right)\frac{b^2}{4\left(\beta^2+\alpha^2\right)}\right]\right.\nonumber\\
&\hspace{0.9in}+\left.\left(C_+-iC_-\right)\sqrt{\beta-i\alpha}~~\Gamma\left[-\frac{1}{2},\left(\beta-i\alpha\right)\frac{b^2}{4\left(\beta^2+\alpha^2\right)}\right]\right\},\nonumber\\
\label{eq:Igeneral}
\end{eqnarray}
where $\Gamma$ is the upper incomplete gamma function.  Note that~(\ref{eq:Igeneral}) may be evaluated in the limit as $b\to0$ to obtain
\begin{equation}
\int_0^\infty\frac{e^{-\beta k^2}}{k^2}\sin(\alpha k^2)~dk=\sqrt{\frac{\pi}{2}}\frac{\alpha}{\sqrt{\sqrt{\alpha^2+\beta^2}+\beta}},~~\beta>0,~~\alpha>0.
\label{eq:Igrowth}
\end{equation}

\subsection{~\label{sec:erfint}}
In this section, general formulas for convergent integrals of the following form are established:
\begin{equation}
I=\int_{-\infty}^\infty\frac{e^{-ak^2}-e^{-ck^2}}{k}e^{bk}~dk,~~a>0,~c>0.
\label{eq:IntA}
\end{equation}
Although we follow a similar procedure as in Appendix~\ref{sec:gammaint}, the variable $b$ has no restriction on its sign, and thus may be used as a differentiation variable in lieu of introducing a new ``dummy'' variable for this purpose.  First,~(\ref{eq:IntA}) is differentiated with respect to $b$ to obtain
\begin{equation}
\frac{dI}{db}=\int_{-\infty}^\infty \left(e^{-ak^2}-e^{-ck^2}\right)e^{bk}~dk.
\label{eq:dIntA}
\end{equation}
Equation~(\ref{eq:IntA}) indicates that the following constraint must be satisfied:
\begin{equation}
I=0~\textrm{ at }~b=0,
\label{eq:IntC}
\end{equation}
since the integrand of~(\ref{eq:IntA}) is odd and $k=0$ is a removable singularity.  By defining 
\begin{equation}
J=\int_{-\infty}^\infty e^{-ak^2}e^{bk}~dk.
\label{eq:IntJ}
\end{equation}
we can solve the first piece of~(\ref{eq:dIntA}) and then combine with the second piece afterwards.  
The exponentials in~(\ref{eq:IntJ}) are rewritten by completing the square, leading to 
\begin{equation}
J=e^{\frac{b^2}{4a}}\int_{-\infty}^\infty e^{-a\left(k-\frac{b}{2a}\right)^2}~dk,
\label{eq:dIntA2}
\end{equation}
and the substitution $u=\sqrt{2a}\left(k-\frac{b}{2a}\right)$ is made to obtain
\begin{equation}
J=\frac{e^{\frac{b^2}{4a}}}{\sqrt{2a}}\int_{-\infty}^\infty e^{-\frac{1}{2}u^2}~du,
\label{eq:dIntA3}
\end{equation}
where the integral in~(\ref{eq:dIntA3}) is exactly equal to $\sqrt{2\pi}$, as defined by the Gamma function. We now have the expression
\[J=\sqrt{\frac{\pi}{a}}e^{\frac{b^2}{4a}},\]
and, after repeating the above analysis for the second piece of~(\ref{eq:dIntA}), we obtain
\[\frac{dI}{db}=\sqrt{\frac{\pi}{a}}\left(e^{\frac{b^2}{4a}}-e^{\frac{b^2}{4c}}\right),\]
which may be integrated with respect to $b$ (applying condition~(\ref{eq:IntC})) to obtain
\begin{equation}
I=\sqrt{\frac{\pi}{a}}\left[\int_0^be^{\frac{b^2}{4a}}~db-\int_0^be^{\frac{b^2}{4c}}~db\right].
\label{eq:IntD}
\end{equation}
After making the variable substitutions $v=\frac{b}{2i\sqrt{a}}$ and $v=\frac{b}{2i\sqrt{c}}$ in the first and second integrals respectively,~(\ref{eq:IntD}) becomes
\begin{equation}
I=2\pi i \left[\int_0^\frac{b}{2i\sqrt{a}}e^{-v^2}~dv-\int_0^\frac{b}{2i\sqrt{c}}e^{-v^2}~dv\right].
\label{eq:IntE}
\end{equation}
The integrals in~(\ref{eq:IntE}) are error functions, thus providing an exact solution for~(\ref{eq:IntA}):
\begin{eqnarray}
I=\int_{-\infty}^\infty\frac{e^{-ak^2}-e^{-ck^2}}{k}e^{bk}~dk=\pi i\left[~\textrm{erf}\left(\frac{b}{2i\sqrt{a}}\right)-\textrm{erf}\left(\frac{b}{2i\sqrt{c}}\right)\right],~~a>0,~c>0~\nonumber\\
\label{eq:IntF}
\end{eqnarray}

\section{Justification for Asymptotic Equivalence in~(\ref{eq:FI}) \label{sec:balance}}
We now justify that the 2$^\textrm{nd}$ term in~(\ref{eq:21d}) is subdominant to the first, which leads to the asymptotic equivalence shown in~(\ref{eq:FI}) as $t\to\infty$.  The typical approach to follow is to rewrite the integral~(\ref{eq:21d}) in two pieces as:
\begin{subequations}
\begin{equation}
h(x,t)={{h}_{1}}(x,t)+{{h}_{2}}(x,t)  
\label{eq:C1a}
\end{equation}
where
\begin{equation}
{{h}_{1}}(x,t)=\frac{1}{2\pi }\int\limits_{-\infty }^{\infty }{{{C}_{1}}}(k){{e}^{{{\phi }_{1}}(k)t}}dk,~~{{h}_{2}}(x,t)=\frac{1}{2\pi }\int\limits_{-\infty }^{\infty }{{{C}_{2}}}(k){{e}^{{{\phi }_{2}}(k)t}}dk 
\label{eq:C1b}
\end{equation}
\label{eq:C1}
\end{subequations}
In~(\ref{eq:C1a}), the path of integration is taken along the real axis as written.  It suffices, then to show that ${{h}_{2}}\ll {{h}_{1}}$ as $t\to \infty $ to justify equation~(\ref{eq:FI}) .  Interestingly, the integral ${{h}_{2}}$ cannot be evaluated via standard techniques in the $t\to \infty $ limit.  In particular the two typical methods at our disposal -- the method of steepest descent and integration by parts -- fail as follows.

\begin{figure*}
\begin{center}
\includegraphics[width=4.5in]{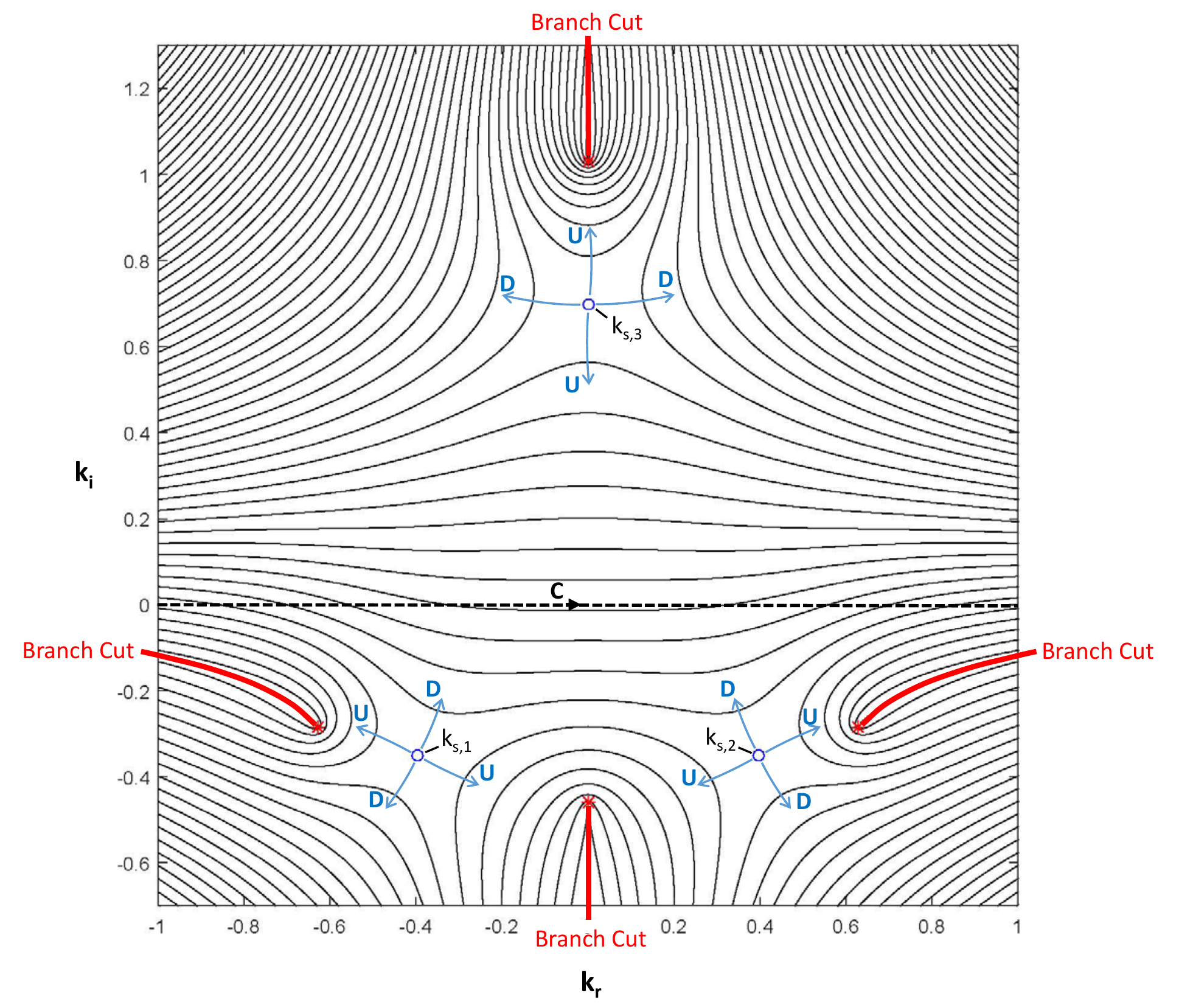}
\end{center}
\caption{Contours of constant $\textrm{Real}[{{\phi }_{2}}(k)]$ vs complex $k={{k}_{r}}+i{{k}_{i}}$ at neutral stability ($R\!e=\cot \theta $) for $x/t=3$, $\theta=\pi/4$, and $W\!e$=0.1.  There are 3 saddle points ($\circ$) denoted as $k_{s,j}$ for $j\in[1,3]$.  Branch points ($\ast$) and branch cuts are taken in accordance with the principle values of the square roots in~(\ref{eq:21a}).   The directions in which $\textrm{Real}[{{\phi }_{2}}(k)]$ decrease and increase are indicated with arrows using the notation D and U, respectively.  The orientation of the saddles does not enable a closed contour to be drawn that includes the contour \textbf{C} (the real axis) and also has $\textrm{Real}[{{\phi }_{2}}(k_{s,j})]$ as a maximum along the contour.  Thus, all the saddle points are spurious, i.e., none may be used via the method of steepest decent to determine the long time asymptotic behavior of the 2$\mathrm{nd}$ term in the integral of~(\ref{eq:21d}).  For reference, the location of the saddles and associated contour values are as follows: $k_{s,1}=-0.2957-0.2492i$, $\textrm{Real}[{{\phi }_{2}}(k_{s,1})]=-2.4192$; $k_{s,2}=0.2957-0.2492i$, $\textrm{Real}[{{\phi }_{2}}(k_{s,2})]=-2.4192$; $k_{s,3}=0.6976i$, $\textrm{Real}[{{\phi }_{2}}(k_{s,3})]=-4.3440$.}
\label{fig:C1}
\end{figure*}

When the method of steepest descent is used, the integral ${{h}_{2}}$ along the real $k$-axis is evaluated as part of a complex contour integral, and the topology of the complex phase function ${{\phi }_{2}}(k)$ is configured such that saddle points, $k_s$, have lower $\textrm{Real}[{{\phi }_{2}}({{k}_{s}})]$ values than surrounding regions of the plane.  Fig.~\ref{fig:C1} shows a typical case that represents the situation for all $x/t$ rays evaluated in this paper.  The indicated saddle points cannot be accessed via a contour integral that includes the real axis (or even a portion of it denoted as the contour C in Fig.~\ref{fig:C1}) in such a way that the saddle points have a maximum Real$[{{\phi }_{2}}({{k}_{s}})]$ along the deformed contour, and the method fails.

The method of integration by parts fails precisely because the integral ${{h}_{2}}(x,t)$ is convergent in the infinite domain of $k$.  This requires its integrand go to zero as $k\to \pm \infty $ (and it does, based on the contours in Fig.~\ref{fig:C1}), and as a result repeated integration by parts merely yields a zero result and no asymptotic behavior can be extracted.  

\begin{figure*}
\begin{center}
\includegraphics[width=4.5in]{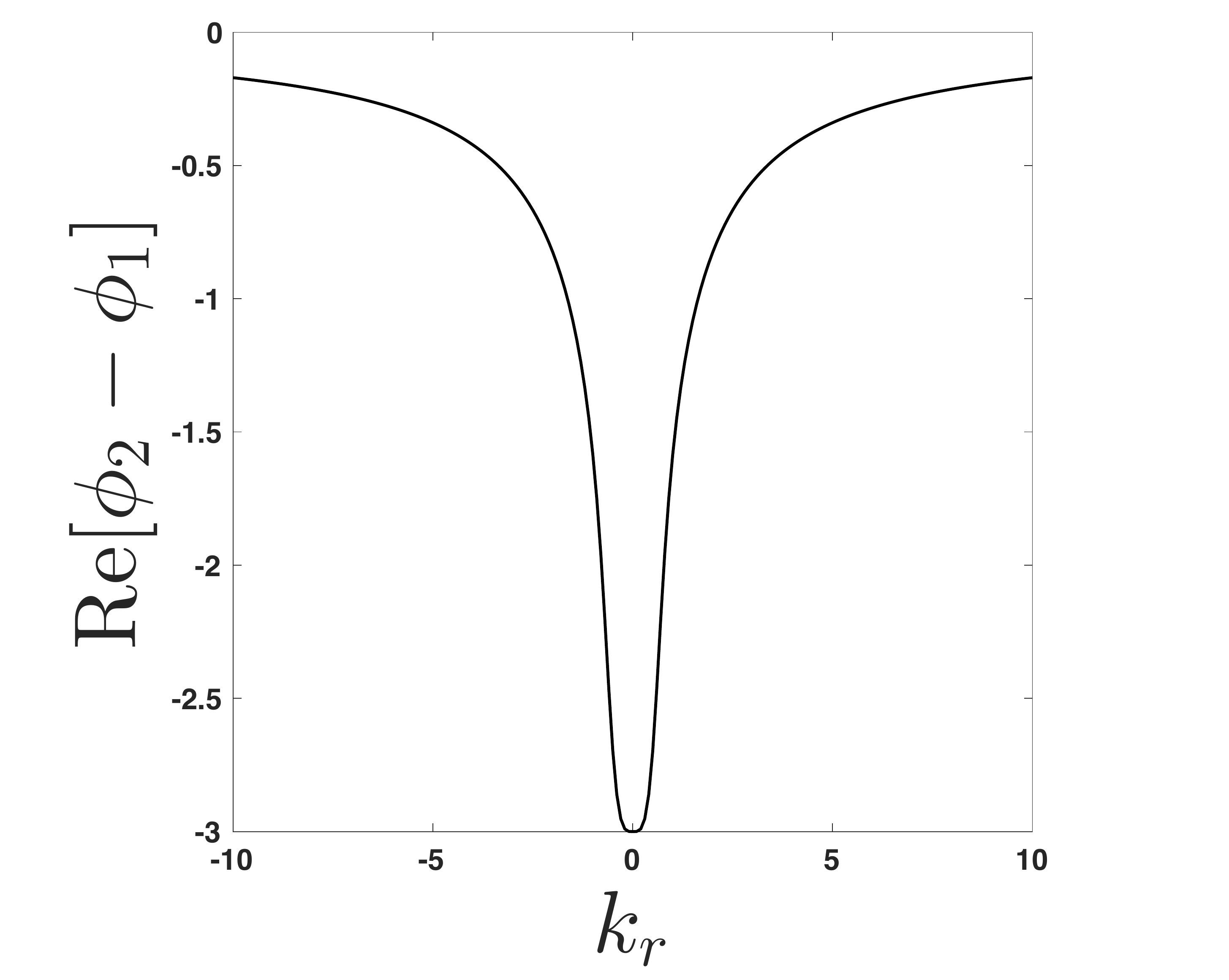}
\end{center}
\caption{Plot of Real$[{{\phi }_{2}}(k)-{{\phi }_{1}}(k)]$ vs. $k$ for $R\!e=\cot(\theta)$, $\theta=\pi/4$, and $W\!e$=0.1.}
\label{fig:C2}
\end{figure*}

A different approach is thus necessary to prove the assertion that ${{h}_{2}}(x,t)$ is subdominant.  To do so, we return to the form of the integral in~(\ref{eq:21d}) and compare the magnitude of the terms in the integrand directly at each value of $k$ (recall that $k$ is real as the path of integration lies along the real axis).  We denote the two pieces of the integrand as:

\begin{subequations}
\begin{equation}
{{I}_{1}}={{C}_{1}}(k){{e}^{{{\phi }_{1}}(k)t}},~~{{I}_{2}}={{C}_{2}}(k){{e}^{{{\phi }_{2}}(k)t}},
\label{eq:C2a}
\end{equation}
and thus using equations~(\ref{eq:21a}),~(\ref{eq:21c}),~(\ref{eq:21e}), we obtain
\begin{eqnarray}
\frac{{{I}_{2}}}{{{I}_{1}}}=\frac{{{C}_{2}}(k)}{{{C}_{1}}(k)}{{e}^{({{\phi }_{2}}(k)-{{\phi }_{1}}(k))t}},~~\frac{{{C}_{2}}(k)}{{{C}_{1}}(k)}=\left( \frac{i{{u}_{0}}-{{\omega }_{1}}{{h}_{0}}}{{{\omega }_{2}}{{h}_{0}}-i{{u}_{0}}} \right),~~{{\phi }_{2}}(k)-{{\phi }_{1}}(k)=-\sqrt{{{\beta }^{2}}-4\gamma }\nonumber\\
\label{eq:C2b}
\end{eqnarray}
\end{subequations}
where all parameters are defined in equations~(\ref{eq:15}) and~(\ref{eq:21}).  As $t\to \infty $ the real part of the exponential in equation~(\ref{eq:C2b}) governs the magnitude of the ratio $I_2/I_1$; a typical plot of the growth rate in the exponential, Real$[{{\phi }_{2}}(k)-{{\phi }_{1}}(k)]$ vs. $k$ is given in Fig.~\ref{fig:C2}.
Fig.~\ref{fig:C2} indicates that as $t\to \infty $, the magnitude of ${{I}_{2}}/{{I}_{1}}\to 0$ for finite $k$.  However, as the domain of $k$ is increased toward $\pm \infty $ (not shown here), the plot asymptotes to zero, indicating that the growth rates are comparable there.  Furthermore, the magnitude of the ratio ${{C}_{2}}(k)/{{C}_{1}}(k)$ in~(\ref{eq:C2b}) approaches 1 in these limits.  Thus, for all finite values of $k$, we can certainly establish that ${{I}_{2}}$ is subdominant to ${{I}_{1}}$ as $t\to \infty $, but this result is not proven in the limit of infinite $k$.  This also indicates that in~(\ref{eq:C1b}), ${{h}_{2}}\ll {{h}_{1}}$ as $t\to \infty $ if finite bounds are taken on the indicated integrals instead of the infinite limits indicated.

To complete the proof that ${{h}_{2}}\ll {{h}_{1}}$ as $t\to \infty $, we rewrite the equations for ${{h}_{1}}$ and ${{h}_{2}}$ in~(\ref{eq:C1b}) as
\begin{subequations}
\begin{equation}
{{h}_{1}}(x,t)=\frac{1}{2\pi }\int\limits_{-\infty }^{-L}{{{C}_{1}}}(k){{e}^{{{\phi }_{1}}(k)t}}dk+\frac{1}{2\pi }\int\limits_{-L}^{L}{{{C}_{1}}}(k){{e}^{{{\phi }_{1}}(k)t}}dk+\frac{1}{2\pi }\int\limits_{L}^{\infty }{{{C}_{1}}}(k){{e}^{{{\phi }_{1}}(k)t}}dk~~
\label{eq:C3a}
\end{equation}
\begin{equation}
{{h}_{2}}(x,t)=\frac{1}{2\pi }\int\limits_{-\infty }^{-L}{{{C}_{2}}}(k){{e}^{{{\phi }_{2}}(k)t}}dk+\frac{1}{2\pi }\int\limits_{-L}^{L}{{{C}_{2}}}(k){{e}^{{{\phi }_{2}}(k)t}}dk+\frac{1}{2\pi }\int\limits_{L}^{\infty }{{{C}_{2}}}(k){{e}^{{{\phi }_{2}}(k)t}}dk~~
\label{eq:C3b}
\end{equation}
\end{subequations}
where $L>0$ is a finite number.  We have already established that the finite bound integral in~(\ref{eq:C3b}) is subdominant to the corresponding finite bound integral in~(\ref{eq:C3a}) from Fig.~\ref{fig:C2} and the preceding arguments (in Fig.~\ref{fig:C2}, $L$=10 as indicated).  Integration by parts may be used on the remaining semi-infinite integrals in equation~(\ref{eq:C3a}) and~(\ref{eq:C3b}) to yield
\begin{subequations}
\begin{eqnarray}
\nonumber
\frac{1}{2\pi }\int\limits_{-\infty }^{-L}{{{C}_{1}}}(k){{e}^{{{\phi }_{1}}(k)t}}dk\sim\frac{{{C}_{1}}(-L)}{{2\pi t~{\left. \frac{d{{\phi }_{1}}(k)}{dk} \right|}_{k=-L}}}{{e}^{{{\phi }_{1}}(-L)t}},\\ \frac{1}{2\pi }\int\limits_{L}^{\infty }{{{C}_{1}}}(k){{e}^{{{\phi }_{1}}(k)t}}dk\sim-\frac{{{C}_{1}}(L)}{2\pi t~{{\left. \frac{d{{\phi }_{1}}(k)}{dk} \right|}_{k=L}}}{{e}^{{{\phi }_{1}}(L)t}},
\label{eq:C4a}
\end{eqnarray}
\begin{eqnarray}
\nonumber
\frac{1}{2\pi }\int\limits_{-\infty }^{-L}{{{C}_{2}}}(k){{e}^{{{\phi }_{2}}(k)t}}dk\sim\frac{{{C}_{2}}(-L)}{{2\pi t~{\left. \frac{d{{\phi }_{2}}(k)}{dk} \right|}_{k=-L}}}{{e}^{{{\phi }_{2}}(-L)t}},\\ \frac{1}{2\pi }\int\limits_{L}^{\infty }{{{C}_{2}}}(k){{e}^{{{\phi }_{2}}(k)t}}dk\sim-\frac{{{C}_{2}}(L)}{{2\pi t~{\left. \frac{d{{\phi }_{2}}(k)}{dk} \right|}_{k=L}}}{{e}^{{{\phi }_{2}}(L)t}}\textrm{ as } t\to \infty.
\label{eq:C4b}
\end{eqnarray}
\end{subequations}
We thus see that 
\begin{eqnarray}
\nonumber
\frac{\frac{1}{2\pi }\int\limits_{-\infty }^{-L}{{{C}_{2}}}(k){{e}^{{{\phi }_{2}}(k)t}}dk}{\frac{1}{2\pi }\int\limits_{-\infty }^{-L}{{{C}_{1}}}(k){{e}^{{{\phi }_{1}}(k)t}}dk}=O({{e}^{({{\phi }_{2}}(-L)-{{\phi }_{1}}(-L))t}}),\\ \frac{\frac{1}{2\pi }\int\limits_{L}^{\infty }{{{C}_{2}}}(k){{e}^{{{\phi }_{2}}(k)t}}dk}{\frac{1}{2\pi }\int\limits_{L}^{\infty }{{{C}_{1}}}(k){{e}^{{{\phi }_{1}}(k)t}}dk}=O({{e}^{({{\phi }_{2}}(L)-{{\phi }_{1}}(L))t}}) \textrm{ as } t\to \infty.
\label{eq:C5} 
\end{eqnarray}

Since according to Fig.~\ref{fig:C2}, which is representative for all cases examined in this paper, Real$[{{\phi }_{2}}(k)-{{\phi }_{1}}(k)]<0$ for finite $k$, this indicates that the ratios in~(\ref{eq:C5}) go to zero as $t\to \infty $.  Thus, we see that all integrals in~(\ref{eq:C3b}) are subdominant to those in~(\ref{eq:C3a}) for all real $k$, which establishes that ${{h}_{2}}\ll {{h}_{1}}$ as $t\to \infty$ in~(\ref{eq:C1}).  This furthermore establishes the asymptotic equivalence indicated in equation~(\ref{eq:FI}) of the main text.  This conclusion is also demonstrated by the agreement between our numerical solutions and the $t\to \infty$ asymptotic behavior of equation~(\ref{eq:FI}).

\section{Long time asymptotic solution to~(\ref{eq:FI})\label{sec:SD}}
The following analysis is separated into two subsections based on a structural change in the dispersion relation for flow down an incline plane.  Classical analysis tells us that, for $R\!e$$\le\cot(\theta)$, the wavenumber of maximum growth is $k=0$ (see Fig.~\ref{fig:cartoon}) and that the exponential growth rate is zero.  Note that, by definition~(\ref{eq:wsaddle}b), a maximum in $\omega_i(k_r)$ is a saddle point.  Any such maximum is also a contributing saddle point, in that a steepest descent path can be closed back to the real-axis (see~\cite{barlow2017}, Appendix A).  Applying definition~(\ref{eq:wsaddle}a) to~(\ref{eq:17}), we find that the corresponding ray of maximum growth is $x/t=3$.  Thus, for $R\!e$$\le\cot(\theta)$, the peak of a wave packet travels at a velocity $x/t$=3 as it flows down the incline plane.  The explicit knowledge of the ($k_s$, $x/t$) pair at the peak allows us to make simplifications outlined above. For $x/t\neq3$, the saddle locations and their corresponding steepest descent paths are less straightforward to deduce.  For this reason, we separate the following evaluation of integral~(\ref{eq:FI}) into two subsections \-- one for the peak along $x/t=3$ and one for the ``off-peak'' rays  $x/t\neq3$. 

\subsection{Evaluation of~(\ref{eq:FI}) for $\frac{x}{t}=3$ and $R\!e\le\cot(\theta)$}
Here, we apply the method of steepest descent to the integral~(\ref{eq:FI}) for $x/t$=3, which corresponds to the contributing saddle $k_s=0$.  Expanding $\phi_1(k)$ (given by~(\ref{eq:21e})) about $k=0$, and making direct use of~(\ref{eq:17}), we obtain the following
\begin{eqnarray}
\nonumber
\phi_1(k)\sim &\left[R\!e-\cot(\theta)\right]k^2-\frac{6}{5}iR\!e \left[R\!e-\cot(\theta)\right]k^3\\ \nonumber&+\left\{\frac{36}{25}R\!e^2 \left[R\!e-\cot(\theta)\right]-\frac{12}{5}\textrm{Re}^3c \left[R\!e-\cot(\theta)\right]^2-\frac{R\!e}{3W\!e}\right\}k^4+O(k^5),~~k\to0,\\
\label{eq:expansion}
\end{eqnarray}
and thus
\begin{subequations}
\begin{equation}
\phi_1(k)\sim \left[R\!e-\cot(\theta)\right]k^2,~~\textrm{ for }R\!e<\cot(\theta)
\end{equation}
\begin{equation}
\phi_1(k)\sim -\frac{\textrm{Re}}{3W\!e}k^4,~~\textrm{ for }R\!e=\cot(\theta),
\end{equation}
\label{eq:leading}
\end{subequations}
which indicates that $k_s=0$ is a 2$^\mathrm{nd}$ order saddle for  $R\!e$$<\cot(\theta)$ and a 4$^\mathrm{th}$ order saddle for $R\!e$$=\cot(\theta)$, according to definition~(\ref{eq:saddle}). Substituting $k=0$ into $C_1(k)$ and~(\ref{eq:leading}) for $\phi_1(k)$ in~(\ref{eq:FI}) leads to
\begin{eqnarray}
\nonumber
h\left( x,t \right)|_{\frac{x}{t}=3}\sim\frac{1}{2\pi}\left(h_0+u_0~R\!e/3\right)\int\limits_{-\infty }^{\infty} e^{-\left[\cot(\theta)-R\!e\right]k^2 t}~dk,~~t\to\infty,~~\textrm{ for }R\!e<\cot(\theta)\\
h\left( x,t \right)|_{\frac{x}{t}=3}\sim\frac{1}{2\pi}\left(h_0+u_0~R\!e/3\right)\int\limits_{-\infty }^{\infty}e^{-\frac{R\!e}{3W\!e}k^4t}~dk,~~t\to\infty,~~\textrm{ for }R\!e=\cot(\theta)\nonumber\\
\label{eq:SD}
\end{eqnarray}
where the integration path remains along the real line, since the argument of the exponential is purely real and thus no rotation through the saddle is required.  Since the integrands of~(\ref{eq:SD}) are even, the integrals may be rewritten
\begin{subequations}
\begin{eqnarray}
h\left( x,t \right)|_{\frac{x}{t}=3}\sim\frac{1}{\pi}\left(h_0+u_0~R\!e/3\right)\int\limits_{0}^{\infty} e^{-\left[\cot(\theta)-R\!e\right]k^2 t}~~dk,~t\to\infty,~~\textrm{ for }R\!e<\cot(\theta)\nonumber\\
\label{eq:SD2a}
\end{eqnarray}
\begin{eqnarray}
h\left( x,t \right)|_{\frac{x}{t}=3}\sim\frac{1}{\pi}\left(h_0+u_0~R\!e/3\right)\int\limits_{0}^{\infty}e^{-\frac{R\!e}{3W\!e}k^4t}~dk,~~t\to\infty,~~\textrm{ for }=R\!e\cot(\theta).\nonumber\\
\label{eq:SD2b}
\end{eqnarray}
\label{eq:SD2}
\end{subequations}
Upon making the variable substitutions $v=\left[\cot(\theta)-R\!e\right]k^2 t$ and $v=\frac{R\!e}{3W\!e}k^4t$ in~(\ref{eq:SD2a}) and~(\ref{eq:SD2b}) respectively, we obtain
\begin{subequations}
\nonumber
\begin{eqnarray}
h\left( x,t \right)|_{\frac{x}{t}=3}\sim\frac{\left(h_0+u_0~R\!e/3\right)}{\pi p\sqrt{\left[\cot(\theta)-R\!e\right]t}}\int\limits_{0}^{\infty} e^{-v}v^{\frac{1}{2}-1}~~dv,~t\to\infty,~~\textrm{ for }R\!e<\cot(\theta)\nonumber\\
\label{eq:SD3a}
\end{eqnarray}
\begin{eqnarray}
h\left( x,t \right)|_{\frac{x}{t}=3}\sim\frac{\left(h_0+u_0~R\!e/3\right)}{\pi p\left\{\frac{\textrm{Re}}{3W\!e}t\right\}^{1/4}}\int\limits_{0}^{\infty} e^{-v}v^{\frac{1}{4}-1}~~dv,~t\to\infty,~~\textrm{ for }R\!e=\cot(\theta)\nonumber\\
\label{eq:SD3b}
\end{eqnarray}
\label{eq:SD3}
\end{subequations}
where the integrals above evaluate to $\Gamma(1/2)$ and $\Gamma(1/4)$ respectively, leading the the results given in~(\ref{eq:asymptoticA}) and~(\ref{eq:asymptoticB}).

\subsection{Evaluation of ~(\ref{eq:FI}) for $\frac{x}{t}\neq3$ and $R\!e\le\cot(\theta)$}
\begin{figure*}
\begin{center}
\includegraphics[width=4.5in]{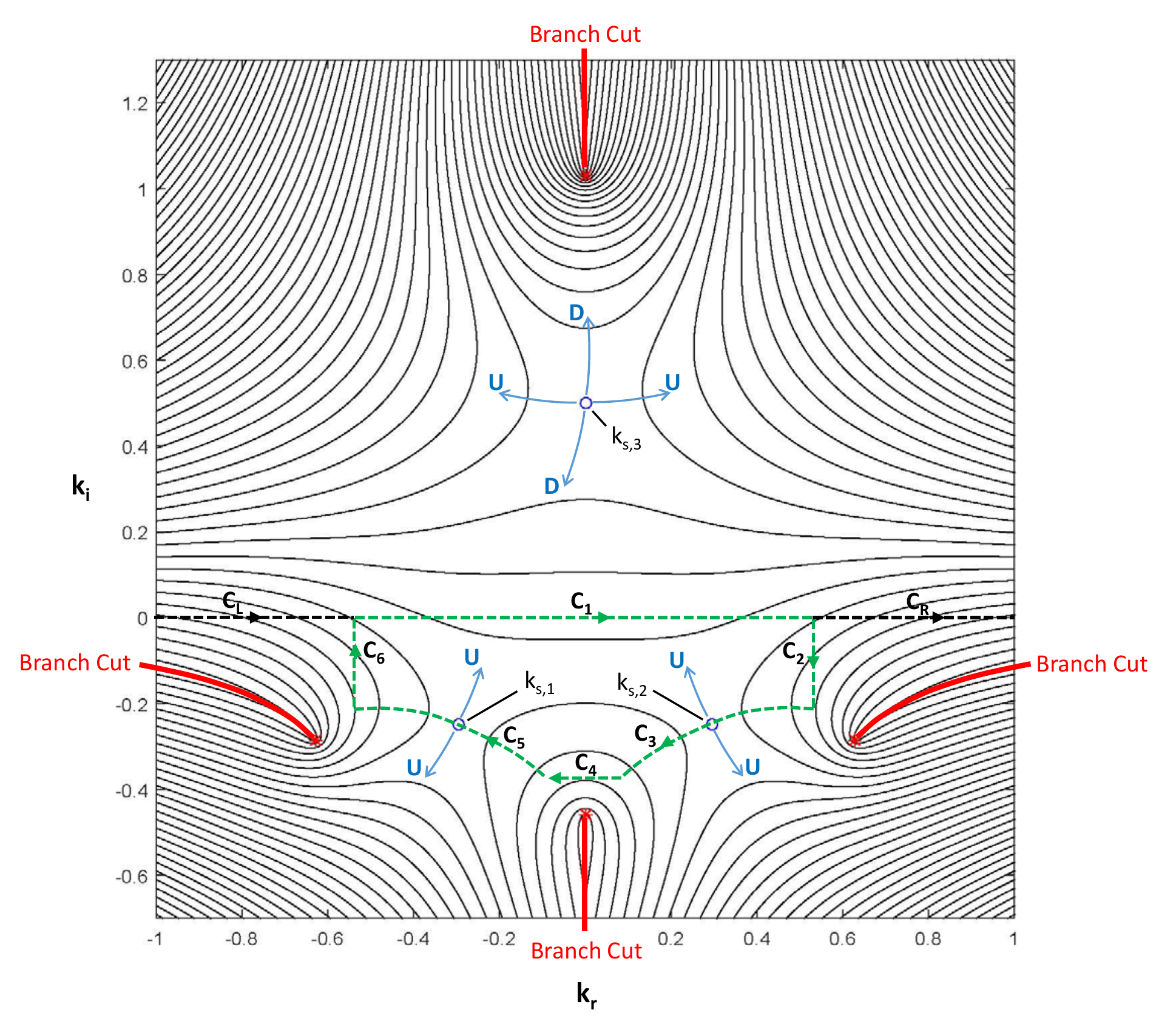}
\end{center}
\caption{Contours of constant $\textrm{Real}[{{\phi }_{1}}(k)]$ vs complex $k={{k}_{r}}+i{{k}_{i}}$ at neutral stability ($R\!e=\cot \theta $) for $x/t=2$, $\theta=\pi/4$, and $W\!e$=0.1. Notation is the same as in Fig.~\ref{fig:C1}.  The orientation of the saddles enables a closed contour (indicated by dashed lines) to be drawn through $k_{s,1}$ and $k_{s,2}$ that includes a portion of the real axis,  \textbf{C$_1$}, and also has $\textrm{Real}[{{\phi }_{1}}(k_{s,j})]$ as a maximum along the contour \--- thus, the method of steepest decent may be used to determine the long time asymptotic behavior of the integral involving $\phi_1(k)$ in~(\ref{eq:21d}) along the contour \textbf{C$_1$}.  Alternatively, the orientation of saddle $k_{s,3}$ makes it spurious (does not contribute to the asymptotic behavior) and is not accessible by connection to any real contour.  The path of integration in~(\ref{eq:21d}) is the \underline{infinite} $k_r$-axis, i.e. consists of the contour \textbf{C}=\textbf{C$_\mathrm{L}\cup$C$_1\cup$C$_\mathrm{R}$}; however, it can be shown via integration by parts that the contributions over \textbf{C$_\mathrm{L}$} and \textbf{C$_\mathrm{R}$} are subdominant to that of \textbf{C$_1$}, such that the long time asymptotic behavior is determined by the integral \textbf{C$_1$} \--- and thus the saddle points dominate the asymptotic behavior.  For reference, the location of the contributing saddles and related quantities necessary to determine the asymptotic behavior using~(\ref{eq:complex}),~(\ref{eq:polar}), and~(\ref{eq:twosaddles}) are as follows: $a$=0.2957, $b$=0.2492, $c$=0.8218, $d$=0.6805, $r$=5.9476, and $q$=5.4227.}
\label{fig:v2}
\end{figure*}

Saddles associated with the off-peak rays $x/t\neq3$ lie in the complex $k$-plane, off of the real-line.  Thus we must take care to evaluate integral~(\ref{eq:FI}) along a closed path that joins the original path (the $k_r$-axis) with a path through the saddle, such that Real[$\phi(k_s)$] is a maximum along this path.   This allows for the integral to be replaced with an approximation near the saddle as $t\to\infty$.  A representative example is given in Fig.~\ref{fig:v2} for $x/t=2$ (parameter values given in the caption). 
Note that there are 3 saddle points associated with $x/t$=2, as indicated by $\bullet$'s in Fig.~\ref{fig:v2}, but only the two lying in the lower half plane are accessible as maxima along a closed path with the real axis.  The steepest descent contour $\int_\textrm{SD}$ is shown by a dashed line.  Note that neither the branch points (indicated by $\ast$'s) nor any other singularities are enclosed between the real axis and the steepest descent path, and so we may apply Cauchy's theorem to obtain
\begin{equation}
h\left( x,t \right)|_{\frac{x}{t}\neq3}\sim\frac{1}{2\pi}\int_\mathrm{-SD}{C}_{1}(k){{e}^{{{\phi }_{1}}(k)t}}dk,~t\to\infty,~R\!e\le\cot(\theta)
\label{eq:Cauchy0}
\end{equation}
where $\int_\textrm{-SD}$ denotes a path from left to right (opposite that shown in Fig.~\ref{fig:v2}) and the long-time asymptotic casting in~(\ref{eq:Cauchy0}) is solely due the omission of the $\phi_2(k)$ term in~(\ref{eq:21}), and not from the path deformation. As Reynolds number, Weber number, $\theta$, and $x/t$ vary, the location of the saddles differ from that shown in Fig.~\ref{fig:v2}.  However, there is always either one or two contributing saddles where Cauchy's theorem may be applied such that~(\ref{eq:Cauchy0}) holds.  Additionally, since all $x/t\neq3$ saddles are 2$^\mathrm{nd}$-order, a general asymptotic solution may be formulated according to the method of steepest descent by looking in the vicinity of the saddles, replacing $C_1(k)$ with $C_1(k_s)$, $\phi(k)$ with its  2$^\mathrm{nd}$-order expansion about $k_s$ in~(\ref{eq:Cauchy0}), and rewriting the limits of integration as follows
\begin{eqnarray}
\nonumber
h\left( x,t \right)|_{\frac{x}{t}\neq3}\sim&\frac{1}{2\pi}\Bigg[{C}_{1}(k_{s1})e^{\phi_1(k_{s1})}\overbrace{\int_{k_{s1}-\epsilon}^{k_{s1}+\epsilon}e^{\frac{1}{2}\left.\frac{d^2\phi_1}{dk^2}\right|_{k_{s1}}(k-k_{s1})^2t}dk}^{I_1}\\ &+~{C}_{1}(k_{s2})e^{\phi_1(k_{s2})}\underbrace{\int_{k_{s2}-\epsilon}^{k_{s2}+\epsilon}e^{\frac{1}{2}\left.\frac{d^2\phi_1}{dk^2}\right|_{k_{s2}}(k-k_{s2})^2t}dk}_{I_2}\Bigg],~t\to\infty,~R\!e\le\cot(\theta)\nonumber\\
\label{eq:Cauchy}
\end{eqnarray}
where $\epsilon$ is a small positive constant. Note that in the above we have assumed two contributing saddles $k_{s1}$ and $k_{s2}$.  For cases with two saddles, such as that shown in Fig.~\ref{fig:v2}, the locations of quantities in~(\ref{eq:Cauchy}) in the complex $k$-plane are as follows
\begin{eqnarray}
\nonumber
k_{s1,2}=\mp a -ib,~\omega_1(k_{s1,2})=\mp c-id,\\
\nonumber
\left.\frac{d^2\phi_1}{dk^2}\right|_{k_{s1,2}}=-r\mp iq,\\
a,~b,~c,~d,~q,~r>0.
\label{eq:complex}
\end{eqnarray}
It is useful to write the following quantity from~(\ref{eq:complex}) in complex polar form
\begin{equation}
r+iq=\rho e^{i\Theta},~\rho=\sqrt{r^2+q^2},~\Theta=\tan^{-1}\left(\frac{q}{r}\right),~\Theta\in\left(0,\frac{\pi}{2}\right)
\label{eq:polar}
\end{equation}
from which it follows from~(\ref{eq:complex}) that
\begin{equation}
\left.\frac{d^2\phi_1}{dk^2}\right|_{k_{s1,2}}=-\rho e^{\pm i\Theta}.
\label{eq:dpolar}
\end{equation}
We now proceed to solve for $I_1$ (denoted in~(\ref{eq:Cauchy})) using the orientation of the path through the saddle $k_{s1}$ shown in Fig.~\ref{fig:v2}.  After substituting~(\ref{eq:dpolar}) into~(\ref{eq:complex}) and splitting $I_1$ into two integrals entering and leaving the saddle $k_{s1}$, we obtain
\begin{equation}
I_1=\int_{k_{s1}-\epsilon}^{k_{s1}}e^{-\frac{1}{2}\rho e^{i\Theta}(k-k_{s1})^2t}dk+\int_{k_{s1}}^{k_{s1}+\epsilon}e^{-\frac{1}{2}\rho e^{i\Theta}(k-k_{s1})^2t}dk.
\label{eq:split}
\end{equation}
The split made above is done such that the quantity $(k-k_{s1})$ can be written in complex polar form in accordance with Fig.~\ref{fig:v2} with the appropriate angle entering and leaving the saddle as follows
\begin{eqnarray}
\nonumber
(k-k_{s1})=Ue^{i\Psi}\\
\nonumber
\Psi\in\left[\frac{\pi}{2},\pi\right]\textrm{  for  }k\in\left[k_{s1}-\epsilon,k_{s1}\right]\\
\Psi\in\left[-\frac{\pi}{2},0\right]\textrm{  for  }k\in\left[k_{s1},k_{s1}+\epsilon\right].
\label{eq:kpolar}
\end{eqnarray}
Upon substituting~(\ref{eq:kpolar}) into~(\ref{eq:split}), we obtain
\begin{equation}
I_1=e^{i\Psi}\int_{-\epsilon e^{-i\Psi}}^{0}e^{-\frac{1}{2}\rho U^2te^{i(\Theta+2\Psi)}}dU+e^{i\Psi}\int_{0}^{\epsilon e^{-i\Psi}}e^{-\frac{1}{2}\rho U^2te^{i(\Theta+2\Psi)}}dU.
\label{eq:U}
\end{equation}
Note that in order for the integrals of~(\ref{eq:U}) to represent paths of steepest descent, the argument of the outer exponential must be real and negative, such that following condition must hold between $\Psi$ and $\Theta$~\cite{bleistein1984}.
\begin{equation}
\Theta+2\Psi=2n\pi
\label{eq:steep}
\end{equation}
where $n=1$ for the first integral of~(\ref{eq:U}) (entering the saddle) and $n=0$ for the second integral of~(\ref{eq:U}) (leaving the saddle).  
In accordance with~(\ref{eq:steep}), we substitute $\Psi=\pi-\Theta/2$ into the first integral of~(\ref{eq:U}) and $\Psi=-\Theta/2$ into the second integral of~(\ref{eq:U}) to obtain (after some simplification)
\begin{equation}
I_1=2e^{-i\frac{\Theta}{2}}\int_0^{\epsilon'}e^{-\frac{1}{2}\rho U^2t}dU.
\label{eq:realint}
\end{equation}
where $\epsilon'=\epsilon e^{i\Theta/2}$ is a real quantity (recall that $U$ is the real-valued magnitude defined in~(\ref{eq:kpolar})).   Thus~(\ref{eq:realint}) is a real integral, translated and rotated from its skewed path through saddle $k_{s1}$ in Fig.~\ref{fig:v2}.  We now replace~(\ref{eq:realint}) with its semi-infinite extension
\begin{equation}
I_1\sim2e^{-i\frac{\Theta}{2}}\int_0^{\infty}e^{-\frac{1}{2}\rho U^2t}dU,~~t\to\infty,
\label{eq:improper0}
\end{equation}
noting that~(\ref{eq:improper0}) asymptotically approaches~(\ref{eq:realint}) as $t\to\infty$, due the to dominant contribution at the saddle. The improper integral in~(\ref{eq:improper0}) evaluates exactly to $\sqrt{2\pi/(\rho t)}/2$ (see \cite{GR} equation 3.321.3), and thus~(\ref{eq:improper0}) becomes
\begin{equation}
I_1\sim e^{-i\frac{\Theta}{2}}\sqrt{\frac{2\pi}{\rho t}},~~t\to\infty.
\label{eq:improper}
\end{equation}
If we apply the same technique as done above to $I_2$ in~(\ref{eq:Cauchy}) involving saddle $k_{s2}$ in Fig.~\ref{fig:v2}, whose real imaginary part is the same but real part opposite sign of $k_{s1}$ and steepest descent path is oriented perpendicular to that of $k_{s1}$, we obtain
\begin{equation}
I_2\sim e^{i\frac{\Theta}{2}}\sqrt{\frac{2\pi}{\rho t}},~~t\to\infty.
\label{eq:improper2}
\end{equation}
Substituting~(\ref{eq:improper}) and~(\ref{eq:improper2}) into~(\ref{eq:Cauchy}), using~(\ref{eq:complex}) to recognize that $\phi_1(k_{s1,2})=\left[b\left(\frac{x}{t}\right)-d\right]\pm i\left[c-a\left(\frac{x}{t}\right)\right]$ and $C_1(k_{s1})=\overline{C_1(k_{s2})}$, we obtain
\begin{subequations}
\begin{eqnarray}
h\left( x,t \right)|_{\frac{x}{t}\neq3}\sim\sqrt{\frac{2}{\rho\pi t}}~e^{\left(b\frac{x}{t}-d\right)t}\textrm{~Real}\left\{C_1(k_{s1})e^{i\left[\left(c-a\frac{x}{t}\right)t-\Theta/2\right]}\right\},~~t\to\infty,~~a\neq0,~~R\!e\le\cot(\theta)\nonumber\\
\label{eq:twosaddles}
\end{eqnarray}
where the $a\neq0$ condition reminds us that we have assumed two contributing saddles symmetric about the imaginary axis, as shown in Fig.~\ref{fig:v2} for saddles in the lower-half plane for $x/t=2$.  The result~(\ref{eq:twosaddles}) may be applied to contributing saddles in the lower or upper half plane (which is the case for $x/t=4$, see Fig.~\ref{fig:v4}) if the definitions in~(\ref{eq:complex}) are modified as follows
\begin{eqnarray}
  a=|\textrm{Real }(k_s)|\\
  b=-\textrm{ Imag }(k_s)\\
  c=|\textrm{Real }[\omega_1(k_s)]|\\
  d=-\textrm{ sgn }[\textrm{Imag }(ks)]~|\textrm{Imag }[\omega_1(k_s)]|\\
  r=\left|\textrm{Imag }\left(\frac{d^2\omega_1}{dk^2}\right)_{k_s}\right|\\
  q=-\textrm{ sgn }[\textrm{Imag }(ks)]\left|\textrm{Real }\left(\frac{d^2\omega_1}{dk^2}\right)_{k_s}\right|.
\label{eq:complex2}
\end{eqnarray}
  \begin{figure*}
\begin{center}
\includegraphics[width=4.5in]{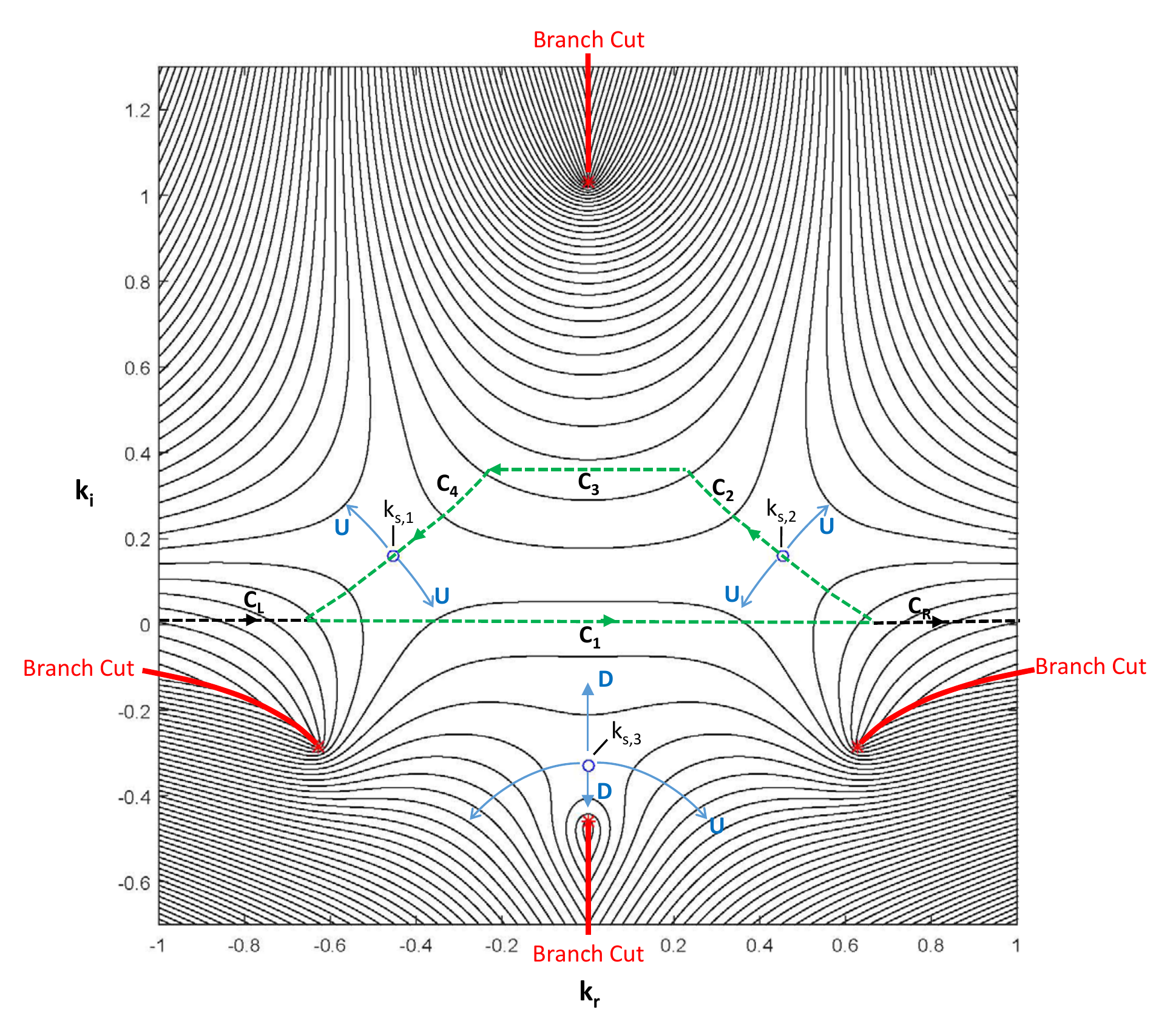}
\end{center}
\caption{Contours of constant $\textrm{Real}[{{\phi }_{1}}(k)]$ vs complex $k={{k}_{r}}+i{{k}_{i}}$ at neutral stability ($R\!e=\cot \theta $) for $x/t=4$, $\theta=\pi/4$, and $W\!e$=0.1. Notation is the same as in Fig.~\ref{fig:C2} and discussion is the same as in Fig.~\ref{fig:v2}.  For reference, the location of the contributing saddles and related quantities necessary to determine the asymptotic behavior using~(\ref{eq:complex}),~(\ref{eq:polar}), and~(\ref{eq:twosaddles}) are as follows: $a$=0.4528, $b$=0.1584, $c$=1.4871, $d$=0.5053, $r$=0.7466, and $q$=5.1297.}
\label{fig:v4}
\end{figure*}
For $R\!e$$<\cot(\theta)$, as either $R\!e$ is decreased or $\frac{x}{t}\to3$ (from above or below) the saddles move closer together until they congeal on the imaginary axis and thus only one saddle contributes, as shown in Fig.~\ref{fig:congeal}.  Carrying out the above analysis for this case leads to
\begin{eqnarray}
h\left( x,t \right)|_{\frac{x}{t}\neq3}\sim\frac{1}{\sqrt{2\rho\pi t}}~e^{\left(b\frac{x}{t}-d\right)t}\textrm{~Real}\left[C_1(k_{s1})e^{i\left(ct-\Theta/2\right)}\right],~~t\to\infty,~~a=0,~~R\!e<\cot(\theta).\nonumber\\
\label{eq:onesaddle}
\end{eqnarray}
\label{eq:offrays2}
\end{subequations}
where $b$, $c$, and $d$ correspond to $k_{s,1}$ in Fig.~\ref{fig:congeal}.
 \begin{figure*}
\begin{center}
\includegraphics[width=4.5in]{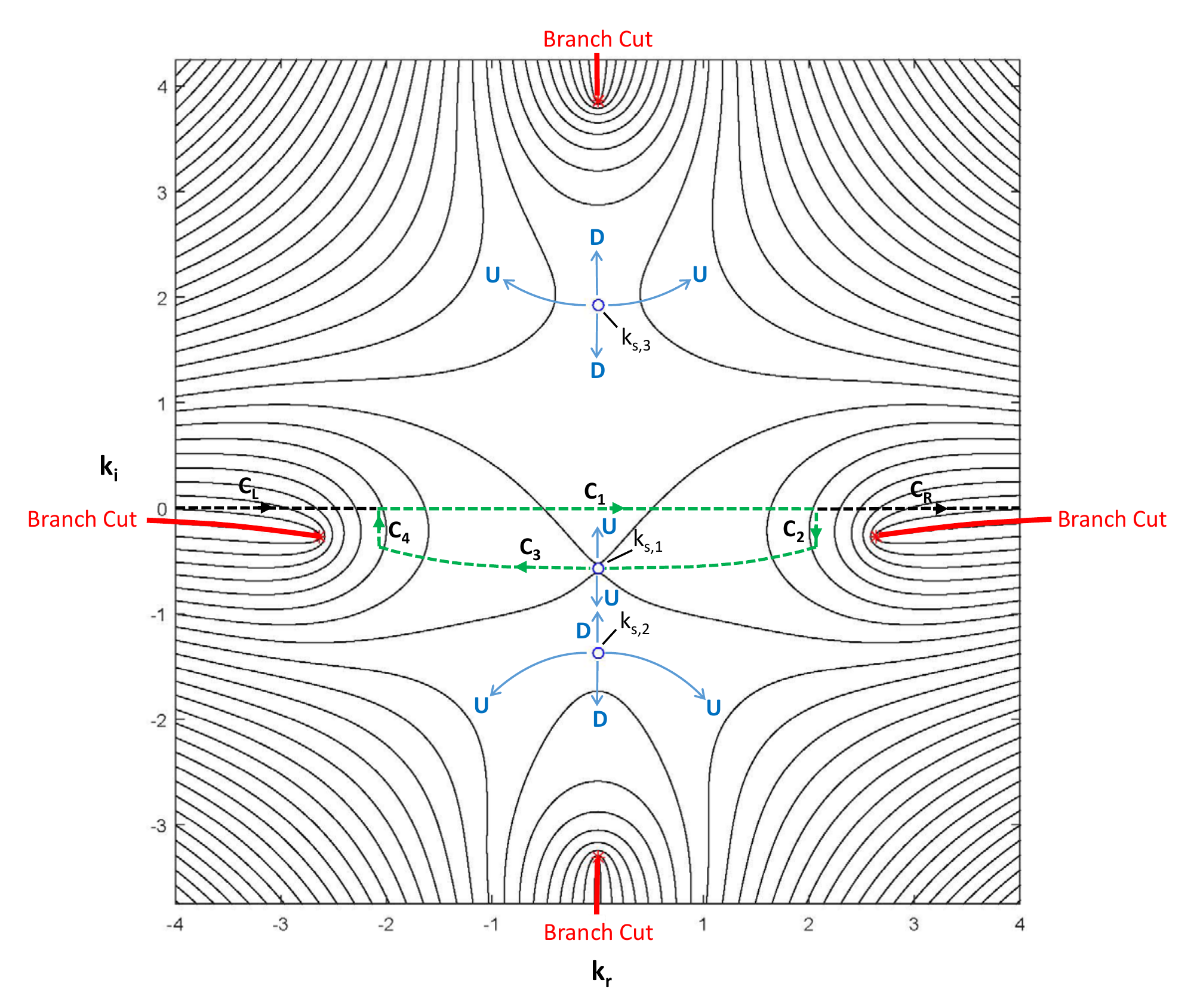}
\end{center}
\caption{Contours of constant $\textrm{Real}[{{\phi }_{1}}(k)]$ vs complex $k={{k}_{r}}+i{{k}_{i}}$ at neutral stability ($R\!e=0.05 \cot \theta $) for $x/t=2$, $\theta=\pi/4$, and $W\!e$=0.1. Notation is the same as in Fig.~\ref{fig:C1}.  The orientation of the saddles enables a closed contour (indicated by dashed lines) to be drawn through $k_{s,1}$ that includes a portion of the real axis,  \textbf{C$_1$}, and also has $\textrm{Real}[{{\phi }_{1}}(k_{s,1})]$ as a maximum along the contour \--- thus, the method of steepest decent may be used to determine the long time asymptotic behavior of the integral involving $\phi_1(k)$ in~(\ref{eq:21d}) along the contour \textbf{C$_1$}.  Alternatively, the orientation of saddles $k_{s,2}$ and $k_{s,3}$ makes them spurious and are not accessible by connection to any real contour.  The path of integration in~(\ref{eq:21d}) is the \underline{infinite} $k_r$-axis, i.e. consists of the contour \textbf{C}=\textbf{C$_\mathrm{L}\cup$C$_1\cup$C$_\mathrm{R}$} and evaluated in the same manner as discussed in Fig.~\ref{fig:v2}.  For reference, the location of the contributing saddle, and related quantities necessary to determine the asymptotic behavior using~(\ref{eq:complex}),~(\ref{eq:polar}), and~(\ref{eq:onesaddle}) are as follows: $a$=0, $b$=1.3728, $c$=0, $d$=2.8420, $r$=1.8563, and $q$=0.}
\label{fig:congeal}
\end{figure*}



\end{document}